# Drug hypersensitivity caused by alteration of the MHC-presented self-peptide repertoire

David A. Ostrov[1], Barry J. Grant[2], Yuri A. Pompeu[3], John Sidney[4], Mikkel Harndahl[5], Scott Southwood[4], Carla Oseroff[4], Shun Lu[1], Jean Jakoncic[6], Cesar Augusto. F. de Oliveira[7], Lun Yang[8], Hu Mei[8], Leming Shi[8], Jeffrey Shabanowitz[9], A. Michelle English[9], Amanda Wriston[9], Andrew Lucas[10], Elizabeth Phillips[10], Simon Mallal[10], Howard Grey[4], Alessandro Sette[4], Donald F. Hunt[9], Soren Buus[5] and Bjoern Peters[4]

1) Department of Pathology, Immunology and Lab Medicine, University of Florida College of Medicine, Gainesville, FL 32611, USA

2) Department of Computational Medicine and Bioinformatics, University of Michigan, Ann Arbor, MI 48109, USA

3) Department of Chemistry, University of Florida, Gainesville, FL, 32611, USA

4) Division of Vaccine Discovery, La Jolla Institute for Allergy and Immunology, 9420 Athena Circle, La Jolla, CA 92037, USA

5) Laboratory of Experimental Immunology, Faculty of Health Sciences, University of Copenhagen, Denmark

6) Brookhaven National Laboratory, Upton, New York, 11973, USA

7) Departments of Chemistry and Biochemistry, Howard Hughes Medical Institute, Center for Theoretical Biological Physics, University of California, San Diego, La Jolla, CA 92037, USA

8) National Center for Toxicological Research, US Food and Drug Administration, 3900 NCTR Road, Jefferson, AR 72079, USA

9) Department of Chemistry, University of Virginia, McCormick Road, Charlottesville, VA 22901, USA

10) Institute for Immunology and Infectious Diseases, Murdoch University, Perth, Western Australia.

**Corresponding authors:**

Bjoern Peters
La Jolla Institute for Allergy and Immunology
9420 Athena Circle
La Jolla, CA 92037, USA
Tel: 858/752-6914
Fax: 858/752-6987
Email: bpeters@liai.org

Howard Grey
La Jolla Institute for Allergy and Immunology
9420 Athena Circle
La Jolla, CA 92037, USA
Tel: 858/752-6568
Fax: 858/752-6987
Email: hgrey@liai.org

## Abstract

Idiosyncratic adverse drug reactions are unpredictable, dose independent and potentially life threatening; this makes them a major factor contributing to the cost and uncertainty of drug development. Clinical data suggest that many such reactions involve immune mechanisms, and genetic association studies have identified strong linkage between drug hypersensitivity reactions to several drugs and specific HLA alleles. One of the strongest such genetic associations found has been for the antiviral drug abacavir, which causes severe adverse reactions exclusively in patients expressing the HLA molecular variant B*57:01. Abacavir adverse reactions were recently shown to be driven by drug-specific activation of cytokine-producing, cytotoxic $CD8^+$ T cells that required HLA-B*57:01 molecules for their function. However, the mechanism by which abacavir induces this pathologic T cell response remains unclear. Here we show that abacavir can bind within the F-pocket of the peptide-binding groove of HLA-B*57:01 thereby altering its specificity. This supports a novel explanation for HLA-linked idiosyncratic adverse drug reactions; namely that drugs can alter the repertoire of self-peptides presented to T cells thus causing the equivalent of an alloreactive T cell response. Indeed, we identified specific self-peptides that are presented only in the presence of abacavir, and that were recognized by T cells of hypersensitive patients. The assays we have established can be applied to test additional compounds with suspected HLA linked hypersensitivities *in vitro.* Where successful, these assays could speed up the discovery and mechanistic understanding of HLA linked hypersensitivities as well as guide the development of safer drugs.

## Introduction

Abacavir is a nucleoside analog that suppresses HIV replication. In about 8% of recipients, abacavir is associated with significant immune mediated drug hypersensitivity, which is strongly associated with the presence of the HLA-B*57:01 allele (1, 2). Three complementary models for the mechanism of immune mediated severe adverse drug reactions have traditionally been discussed(3, 4): *The hapten (or pro-hapten) model* states that drugs and their metabolites are too small to be immunogenic on their own, but rather act like haptens and modify certain self-proteins in the host that lead to immune recognition of the resulting hapten:self-peptide complexes as *de novo* antigens(5-7); *The p-i model* (short for **p**harmacological interaction with **i**mmune receptors) states that drugs can induce the formation of HLA:drug complexes that can activate T cell immune responses directly without requiring a specific peptide ligand (8); *The danger model*, which is in principle compatible with other models, states that danger signals other than the drug itself (such as chemical, physical or viral stress) are required to overcome immune tolerance barriers that otherwise suppress drug hypersensitivity reactions (7).

None of these existing models provide a convincing mechanism for how abacavir induces adverse reactions through the activation of $CD8^+$ cells in a strictly HLA-B*57:01 restricted manner, as was described in a ground breaking paper by the McCluskey group (2). For the hapten hypothesis to apply, abacavir would need to modify one or more self-ligands that are solely presented by HLA-B*57:01 **(Figure 1)**; an unlikely proposition given that HLA molecules are known to fall in groups of overlapping binding specificity, and HLA-B*57:01 has a similar binding motif to the abacavir insensitive HLA-B*58:01 (9). Also, no natural HLA specific drug haptenated peptide has been identified to date, although this has been attempted at least for carbamazepine(10, 11). For the p-i model to apply, abacavir would need to bind to a unique surface patch of HLA-B*57:01 that is capable of inducing TCR recognition. However, the two residues that distinguish abacavir sensitive HLA-B*57:01 from insensitive HLA-B*57:03 are located at the bottom of the HLA binding groove and are unlikely to contact the T cell receptor. Finally, the danger model may well be relevant for abacavir adverse reactions, but does not explain its HLA restriction. We find an alternative hypothesis more attractive, namely that the binding groove of HLA-B*57:01 can accommodate abacavir(12), thereby altering the repertoire of self-peptide ligands that is bound and presented(2, 10). This could lead to a primary and polyclonal immune response, which is in line with observations that abacavir can induce a relatively diverse response in T cells from abacavir naïve individuals after *in vitro* stimulation for 11 days(2) with broad usage of V beta receptors by the responding T cells(2). However, to date there has been no experimental evidence that supports this 'altered self repertoire' hypothesis as a mechanism for drug hypersensitivity.

## Results

To determine whether abacavir can impact the peptide-binding specificity of HLA-B*57:01, we tested its binding affinity using positional scanning combinatorial peptide libraries(13) in the presence and absence of the drug. Each library consisted of 9-mer peptides that shared the same residue at one position, but were otherwise random in sequence. In the absence of abacavir, our binding measurements reproduced the known motif of HLA-B*57:01 for C-terminal peptide residues, namely a preference for large hydrophobic residues such as tryptophan and phenylalanine, and disfavoring small hydrophobic residues like alanine and valine (**Figure 2A**). We were specifically interested in residues showing an increased affinity in the presence of abacavir, which could potentially lead to presentation of de novo peptides. The most dramatic gain was found for peptides with C-terminal valine (8.8 fold increase), alanine (6.7 fold increase), and isoleucine (5.5 fold increase). The only other residue with a 5-fold or higher increase was leucine at position 7 (**SI Appendix, Table S1**). In contrast, the control MHC molecule HLA-B*58:01 showed no increases in affinity above a factor of 3 for any residue at any position **(Figure 2B and SI Appendix, Table S2).**

Based on these results we synthesized individual peptides with the sequence HSITYLLPV (pep-V) and HSITYLLPW (pep-W). Residues 1-8 of these peptides were chosen based on their high affinity in combinatorial library scans in the presence of abacavir. The C-terminus was chosen so that pep-V was expected to be unable to bind efficiently in the absence of abacavir, but would do so in its presence. As a control, pep-W with a C-terminal tryptophan was expected to be able to bind HLA-B*57:01 readily in the presence or absence of abacavir. The peptides were radiolabeled and tested for binding in increasing concentrations of abacavir. Peptide binding assays demonstrate that pep-V requires abacavir in a dose-dependent manner to bind a detectable amount of peptide (**Figure 2C**), while pep-W bound well regardless of the presence or absence of abacavir (**Figure 2D**). Similar results were obtained using two independent peptide-HLA-B*57:01 interaction assays measuring either binding (14) or stability (15) (**SI Appendix, Figure S1**) and an additional control allele, HLA-B*57:03 **(SI Appendix, Table S3)**. In summary, we found that specific peptides such as pep-V have a significantly increased affinity for HLA-B*57:01 in the presence of abacavir, that this effect is abolished when switching the C-terminal P9 residue to a tryptophan (pep-W), and that this effect is not observed for control HLA alleles.

Structural analysis was used to further dissect the mechanism by which abacavir may facilitate the binding of pep-V to HLA-B*57:01. Computational solvent mapping(16), molecular docking(17), and molecular dynamics simulations(18) of 30ns in length identified a potential binding site for abacavir localized to the F-pocket in the vicinity of residue Ser 116 that was shown to be required for abacavir T cell recognition(2) **(SI Appendix, Figure S2)**. In contrast, a similar protocol did not identify stable complexes of MHC, abacavir and peptide with either pep-W or abacavir insensitive HLA-B*57:03.

To directly test our hypothesis that abacavir binds within the antigen-binding cleft, we solved the X-ray crystal structure of HLA-B*57:01 bound to pep-V in the presence of abacavir. We believe that choosing pep-V instead of a ligand with a large side chain at the C-terminus such as those used in previous crystal structures(2) was crucial for obtaining crystals that could resolve the location of abacavir. The structure was refined to an R value of 18 % and $R_{free}$ of 22 % using X-ray diffraction data to 2.0 Å (PDB ID 3UPR; see Methods for details and **SI Appendix, Table S4** for full refinement statistics). Abacavir is bound to a largely hydrophobic pocket in the antigen-binding cleft forming van der Waals contacts with both HLA-B*57:01 (Tyr9, Tyr 74, Ile 95, Val97, Tyr99, Tyr123, Ile 124, Trp147) and pep-V (HSITYLLPV) (Ile3, Leu7, Val9) **(Figure 3A and SI Appendix, Figure S3)**. To determine if the conformation of pep-V is altered by abacavir binding compared to conventionally presented peptides, we compared our structure to four published structures of peptides bound to HLA-B molecules. As shown in **Figure 3D**, the main chain conformation of pep-V is similar to other peptides bound to HLA-B. The amino- and carboxy- termini of pep-V and the other HLA-B bound peptides shown in **Figure 3D** are buried in the A and F pockets, respectively, forming conventional contacts with highly conserved residues. The central portion of the pep-V main chain is within the range of variability demonstrated by other peptide/HLA-B complexes. The pep-V main chain does not protrude in a central bulge as does the longer 11-mer peptide bound to HLA-B*35:01. These data suggested that pep-V in the presence of abacavir is bound in a regular antigen conformation allowing for conventional recognition by TCRs rather than requiring hapten or superantigen recognition modes(19, 20).

Only two residues, Asp114 and Ser116, distinguish HLA-B*57:01 from the abacavir insensitive allele HLA-B*57:03. Abacavir interacts directly with these residues **(Figures 3C and 3E)**. The Oδ1 atom of Asp114 is within H bonding distance of the main purine group N2 and N3 atoms of abacavir. The hydroxyl group of Ser116 forms an H bond with the 2-amino group on the purine ring of abacavir. The exchange of Ser116 to Tyr116 which is found in HLA-B*57:03 is expected to disrupt these interactions. Indeed, it was shown that this single residue exchange is sufficient to abrogate abacavir associated recognition by $CD8^+$ T cells(2). The HLA-B*58:01 allelic variant is also very similar to HLA-B*57:01. These molecules have identical amino acids at positions 114 and 116, but differ at five other positions. These include Val97 which is part of the hydrophobic pocket in HLA-B*57:01 that forms van der Waals contacts with abacavir (**SI Appendix, Figure S3**), and is replaced with a charged Arg97 in HLA-B*58:01 abrogating these interactions. In addition, the structure reveals that the side chain of Val9 of pep-V is within van der Waals contact distance of the cyclopropyl moiety of abacavir. Finally, the contact made between the Leu7 residue of the peptide and abacavir explains why the MHC binding assays showed that this residue had the highest increase in affinity in the presence of abacavir apart from C-terminal residues. In summary, our findings give a structural explanation for why distinct repertoires of peptides with short hydrophobic P9 side chains are bound by HLA-B*57:01 in the presence of abacavir, whereas other HLA alleles are unaffected.

To explore the biologic relevance of these findings, we determined whether live cells treated with abacavir present a different set of self-peptides on HLA-B*57:01 molecules than untreated cells. Our binding assays predicted that HLA-B*57:01 in the presence of abacavir would favor presentation of peptides having a small C-terminal residue such as valine and isoleucine rather than tryptophan and phenylanine normally preferred by HLA-B*57:01 molecules. To answer this question, peptides were eluted from a HLA-B*57:01 single allele transfected 721.221 cell line (21) treated with and without abacavir. Eluted peptides were analyzed by nanoflow-HPLC coupled to an Orbitrap mass spectrometer equipped with a front-end electron transfer dissociation (FETD) (22, 23). We identified 539 and 682 peptide sequences from the drug treated and untreated samples, respectively, 287 of which were found in both samples (**SI Appendix, Table S5**). No peptides with valine at the C-terminus were identified in untreated cells, but fifteen peptides with valine at the C-terminus were identified in the presence of abacavir. Three of these peptides were present at levels (>100 copies/cell) that place them among the top 5% of all peptides in the drug treated sample. **Table 1** compares the frequency of C-terminal residues in peptides identified uniquely in either the abacavir treated or untreated samples. In the presence of abacavir, we found not only a significant enrichment for peptides with valine at the C-terminus, but also for isoleucine. In contrast, there were significantly fewer peptides with tryptophan and phenylalanine at the C-terminus. The results for these residues exactly matched the predictions made by the binding assays. In contrast to valine, no peptides with C-terminal alanine were discovered in the presence of abacavir, even though both showed a similar increase in affinity. However, the absence of peptides with an alanine at the C-terminus can be explained by the antigen processing machinery including proteasomal cleavage and TAP transport, which restricts the peptide repertoire available for binding to MHC, and disfavors peptides with C-terminal alanine (24, 25). In summary, we find that the self-peptide repertoire presented by HLA-B*57:01 positive cells in the presence of abacavir is significantly altered in a manner that is consistent with results obtained from the molecular MHC binding assays.

Finally, we set out to determine if we could detect T cells in hypersensitive patients that recognize HLA B*57:01 restricted peptides that are presented only in the presence of abacavir. Of note, we did not expect high frequency T cell responses against any individual peptide, as the altered ligand mechanism suggested that the response was directed against a very large number of different peptides. We screened PBMCs from five HLA-B*57:01 positive donors with a clinical history of abacavir hypersensitivity for recognition of peptides with valine at the C-terminus that were identified following elution from abacavir treated HLA-B*57:01 cells. PBMCs were incubated for 15 minutes with a high concentration of 10 µg/ml endogenous peptides in the presence of 100 µg/ml abacavir followed by washing to optimize loading of HLA B*57:01 with the specific exogenously added peptide(s) and reduce abacavir entering cells and enabling presentation of other endogenous ligands. After a first round of screening four pools of 3-4 peptides in ELISPOT assays using PBMCs from two donors with sufficient samples available, the peptides from the pool with the highest response (pool 1) were tested individually. Peptide VTTDIQVKV gave the highest response in this screen (**SI Appendix, Figure S4**). Subsequently, we tested the four pools and the individual peptide VTTDIQVKV in the remaining three donors. As shown in Figure 4, a significantly higher response was detected when cells were pulsed with peptide VTTDIQVKV and abacavir compared to the response against cells pulsed with either abacavir alone or peptide alone. These data demonstrated that memory T cell responses in abacavir hypersensitive donors were directed against a self-peptide that required abacavir to efficiently bind HLA B*57:01.

## Discussion

Our findings provide a mechanistic explanation for abacavir-induced adverse drug reactions. We found that abacavir can bind inside the peptide binding groove of HLA-B*57:01, thereby enabling the presentation of peptide ligands that normally cannot bind in substantial amounts. Since T cells are generally tolerant only to MHC-restricted peptide ligands presented during T cell development in the thymus (26), presentation of an altered repertoire of class I MHC binding peptides will be perceived as being foreign and trigger $CD8^+$ T cell responses. These responses have been previously demonstrated to be a hallmark of patients with abacavir hypersensitivity (2). Although abacavir is metabolized inside cells(27, 28), our finding that HLA-B*57:01 is affected by the abacavir parent drug suggests that the parent drug itself is present during peptide loading in the ER. Indeed, others have detected unaltered intracellular abacavir which rapidly co-localizes with HLA-B57 in the endoplasmic reticulum (27, 29). Abacavir might well be delivered to the ER directly from the extracellular medium without traversing the cytosol, as was described for HLA class I binding peptides (30). In this context, it should also be noted that only a small fraction of peptide:MHC loading events have to be affected by abacavir to cause a physiological effect as T cells have a very high sensitivity to detect non-self peptides which is essential to their function, and remarkably few peptide:MHC complexes are sufficient to trigger a T cell response(31, 32).

These biochemical and structural findings provide an explanation for why, of the more than 5000 Class I MHC alleles, only HLA-B*5701 is affected by abacavir. This also provides support for the altered peptide model where the drug in effect creates a novel HLA-B allele presenting self-peptides to which the host has not been toleralized, analogous to the situation occurring in mismatched HLA organ transplantation. In organ transplantation, it has recently been reported that pre-existing Class I restricted effector memory T-cell responses to prevalent viral infections can mediate organ rejection (33). This model of heterologous immunity may explain the clinical manifestations that arise from drug-induced altered peptide presentation, and why only 55% of HLA-B*5701 positive patients treated with abacavir develop hypersensitivity. Other explanations for the incomplete positive predictive value of specific HLA alleles for abacavir and other drug hypersensitivity syndromes and the varying clinical features is that the relevant peptide recognized by drug specific T-cells in the presence of drug is itself genetically polymorphic and/or only present in some patients or tissues or that only some patients have a T-cell clonotype able to respond to the neo-antigen(34).

Over the last 10 years in particular there have been many reported HLA-associated drug toxicities(4). Further studies should explore in detail if the mechanism for HLA-linked adverse reactions to abacavir applies to adverse reactions against other small molecules that seem to be immune mediated and HLA linked, such as chronic beryllium disease(35) adverse reactions to allopurinol(36) or carbamazepine(37). Our methods and findings are particularly significant for such studies as they can be utilized to identify the structural, biochemical and functional basis of potential HLA associated T-cell mediated drug hypersensitivities before use of a drug in man. This may have utility both in excluding high risk compounds from further development, and guiding the design of compounds that do not bind risk HLA alleles or alter the repertoire of peptides presented. The biochemical and functional assays described could also be used to characterize the HLA restriction and likely immunopathogenesis of cases of hypersensitivity in early clinical studies. This could facilitate the early introduction of HLA screening as has been successfully implemented for the prevention of abacavir hypersensitivity (38), rather than having to rely exclusively on genetic association studies that require large cohorts of affected patients. Finally, these findings have potential relevance for a broader understanding of HLA associations in the immunopathogenesis of autoimmune diseases, infectious diseases and cancer. Such associations can be strong, yet remain enigmatic, because the molecular recognition events underlying the associations are unclear. The discovery that a small-molecule drug can bind within the antigen-binding cleft of MHC and alter the repertoire of presented peptides suggests new mechanisms of action as the causative basis for HLA associations. HLA-associated disorders may be perpetuated by drugs, small molecules of environmental origins or self-metabolites that bind within the antigen-binding cleft and alter peptide binding.

## Materials and Methods

**X-ray crystallography.** Refolded $\beta_2$-microglobulin, HLA-B*57:01 and abacavir formed crystals at 4 mg per ml in 0.17 M sodium acetate trihydrate, 0.1 M sodium cacodylate, pH 6.5, 25% PEG 8,000, 15% glycerol which belong to the space group $P2_1$ with unit cell dimensions a=44.8 Å, b=130.7 Å, c=88.3 Å, α=90°, β=104.6°, γ=90°. Two peptide/abacavir/HLA complexes in the asymmetric unit were identified by molecular replacement using coordinates for the heavy chain and $\beta_2$-microglobulin from PDB 2RFX (HLA-B*57:01 bound to LSSPVTKSF). The peptide and ligands were not included in the molecular replacement model. Unambiguous electron density for the peptide and abacavir were visible in Fo-Fc difference maps and simulated annealing omit electron density maps. The structure was refined to an R value of 18% and $R_{free}$ of 21% using X-ray diffraction data to 2.0 Å.

The best crystals diffracted to a high-resolution limit of 1.9 Å. The crystals belonged to space group $P2_1$ and contained two HLA heterodimers (heavy chain HLA-B*57:01 and light chain $\beta_2$-microglobulin) in the asymmetric unit. The phasing was done by molecular replacement and the initially calculated Fo-Fc difference map showed unambiguous electron density for both the peptide and the abacavir drug.

**Abundance calculation for eluted peptides.** Peptide abundances were determined by comparing the ion current observed for 100 fmol of an internal standard, angiotensin I (DRVYIHPFHL), to that observed for individual peptides. This femtomole quantity was then converted to peptide molecules by multiplying by Avogadro's number and to molecules (copies)/cell by dividing this number by the total number of cells employed to generate the sample aliquot injected into the mass spectrometer.

**T cell ELISPOT assays.** Clinical samples: Informed consent was obtained from all patients and studies approved by both the Royal Perth Hospital and Murdoch University ethics committees, Perth, Western Australia Cryopreserved PBMC from abacavir hypersensitive patients of time-points previously evaluated by abacavir specific ELISpot assay were thawed and cultured O/N in RPMI 1640 media containing 10% fetal calf serum (FCS), 50 U/ml Penicillin, 50 µg/ml streptomycin and 1mM sodium pyruvate O/N (F10 medium)(Life Technologies). The assays were controlled using a unstimulated PBMC control, a positive control for abacavir induced endogenous ligand consisting of 10 µg/ml of abacavir added for the length of the assay culture, an abacavir 15 minute pulse and wash control to control for effects of the pulse on inducing endogenous peptide targets. Peptide +/- abacavir incubation times were derived to achieve the best balance between detection of enhanced exogenous peptide binding and a restriction of abacavir induced endogenous peptide presentation. To setup the assay, $5 \times 10^5$ PBMC, were then exposed to 10 µg/ml exogenous peptides, either singly or as peptide pools containing up to 4 peptides, in the presence of absence of 100 µg/ml abacavir for 15 minutes within a 37°C cell culture incubator. The cells were then immediately washed with 25 volumes of ice cold RPMI 1640 cell culture media and centrifuged for 10 minutes at 300g at 4°C. Cells were then re-suspended in 450 µl of F10 medium containing 20U/ml rh IL-2 (Peprotech). $2 \times 10^5$ PBMC in duplicate were then transferred into MAIPS4510 ELISpot plates coated with 2 µg/ml anti-IFN-γ antibody (1-D1k, Mabtech) and blocked with F10 medium. Cells were cultured O/N in a 37°C cell culture incubator. The following day the plates were washed with sterile PBS and incubated with anti-IFN-γ biotinylated antibody (7-B6-1, Mabtech) for 2 hours, washed and incubated for 1 hour with streptavidin-HRP and then washed. Plates were substrated for 8-12 minutes using TMB substrate (Mabtech). Plates were air dried and then counts evaluated using an AID automated microplate ELISpot Reader (AID GmbH. Strassberg, Germany). The sequences of the peptides tested were as follows, Pool 1: KTIHLTLKV, RTLAEIAKV, VTTDIQVKV, TVAPFNPTV. Pool 2: HSIPVTVEV, KSNGTIIHV, RTFHHGVRV. Pool 3: ATIKLQSTV, KIYEGQVEV, RSARVTVAV, RVAGIHKKV. Pool 4: RSVALAVLA, KAAKIRVSV, KVAKVEPAV, RTTETQVLV.

## Acknowledgements

We want to thank Amiyah Steen, Sandy Ngo, Carrie Moore and Victoria Tripple for technical assistance, Patrick Hogan for helpful discussions, and Janet Woodcock and Donna Mendrick for support. Funding was provided by NIH grant AI 33993 to D.F.H and HHSN 272 200900045C to S.B. The views presented in this article are those of the authors and do not necessarily reflect the views or policies of the US Food and Drug Administration.

**Table 1: Distribution of C-terminal residues in peptides uniquely presented by abacavir treated and untreated cells**

| C-terminal residue | abacavir untreated | abacavir treated | p-value *) |
|:---:|:---:|:---:|:---:|
| **W** | **218** | **↓ 95** | **2E-05** |
| **F** | **89** | **↓ 31** | **1E-03** |
| Y | 42 | 33 | 0.38 |
| L | 25 | 25 | 0.10 |
| **I** | **14** | **↑ 45** | **1E-09** |
| M | 6 | 7 | 0.39 |
| **V** | **0** | **↑15** | **5E-07** |

*) Two-tailed Fisher exact test. All C-terminal residues for which two or more peptides were identified are listed

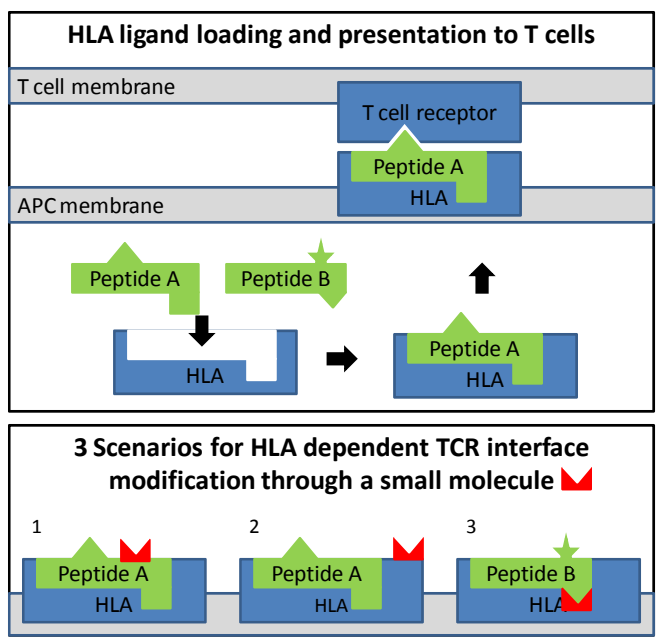

**Figure 1. Schematic presentation of HLA antigen presentation and HLA linked mechanisms of adverse reactions. Top panel:** T cell receptors monitor the universe of antigens an individual is exposed to by surveying the ligands presented on an antigen presenting cell (APC) membrane in the context of HLA molecules. The HLA ligands are typically peptides loaded onto the HLA molecule inside the APC and subsequently exposed on the surface. Different allelic variants of HLA molecules have different binding specificities, resulting in a specific profile of presented ligands. In the example shown, peptide A but not peptide B can bind to the HLA molecule. Self-peptides presented to T cells in this manner do not trigger an immune response, as T cells that are self-reactive are negatively selected during thymic development. However, when T cells encounter an unknown ligand (for example a virus derived peptide), an immune response is triggered. **Bottom panel:** There are three scenarios for HLA dependent drug induced modifications that affect the TCR interface. 1) A ligand that is uniquely presented by the HLA allele is modified by the drug. 2) The HLA molecule itself is modified in a region exposed to the TCR and 3) The binding specificity of the HLA molecule is altered by the presence of the drug, resulting in presentation of novel ligands such as peptide B.

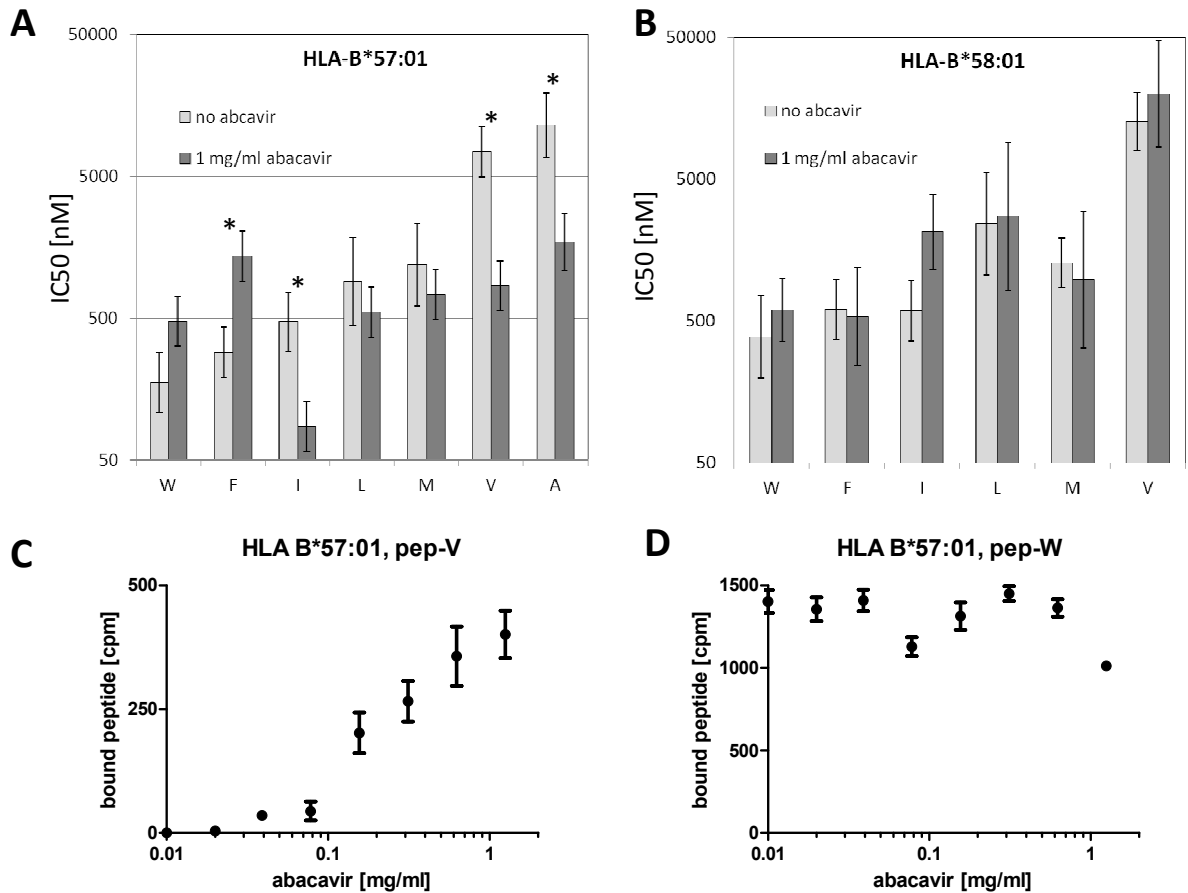

**Figure 2: The presence of abacavir alters the binding specificity of HLA-B*57:01.** Combinatorial peptide libraries were tested for binding to HLA-B*57:01 **(A)** and HLA-B*58:01 **(B)** in competitive binding assays described in (13, 39). Results for libraries with different C-terminal residues are shown for those residues with affinities of 5,000 nM or better - a minimal threshold for binding. Error bars indicate 95% confidence intervals for the mean, and residues marked with an asterisk * had significantly different IC50 values in the presence vs. absence of abacavir (p < 0.001, two-tailed Student's t-test comparing log(IC50) values). The most pronounced affinity increases for HLA-B*57:01 in the presence of 1mg/ml abacavir were found for peptides with a valine at the C-terminus, which increase more than 8-fold in affinity, followed by alanine and isoleucine (all above 5-fold). In contrast, the maximum affinity increase for any peptide library binding to HLA-B*58:01 was less than 3-fold. **(C,D)** Individual Peptides HSITYLLPV (pep-V) and HSITYLLPW (pep-W) were radiolabeled and tested for binding to HLA-B*57:01 and B*58:01 in increasing doses of abacavir. After washing, no pep-V binding to HLA-B*57:01 was detectable in the absence of abacavir, but strong binding was detected in its presence. No significant effect of abacavir was observed for pep-W binding to HLA-B*57:01.

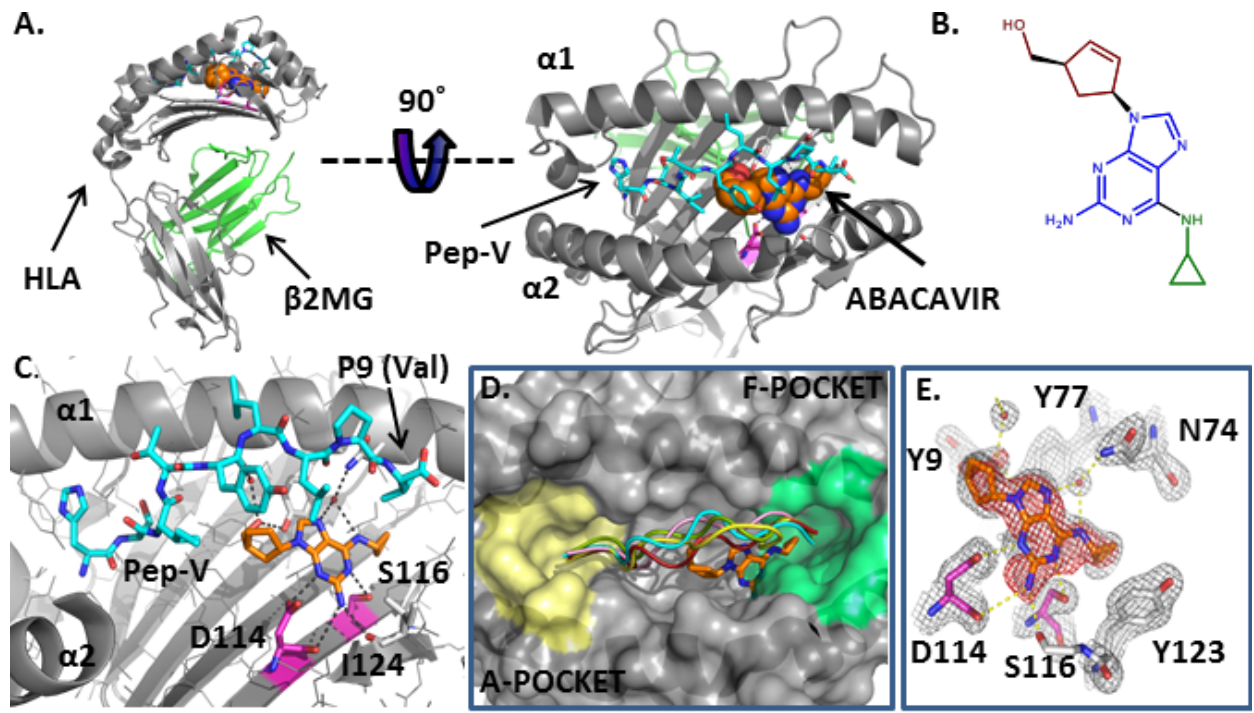

**Figure 3: The crystal structure of the abacavir:peptide:MHC complex solved to a resolution limit of 2.0 Å reveals intermolecular contacts within the antigen-binding cleft.** Panel **A** shows a cartoon diagram of HLA-B*57:01 in gray. The peptide HSITYLLPV is shown in cyan carbons. Abacavir is shown as spheres, orange for carbon, blue for nitrogen and red for oxygen. Panel **B** shows the chemical structure of abacavir, with the cyclopropyl moiety colored in green, the purine core in blue and the hydroxymethyl cyclopentene moiety in red. Panel **C** shows that abacavir forms H bond interactions (black dashes) with both the peptide and HLA-B*57:01. The residues that distinguish the abacavir sensitive allele HLA-B*57:01 from abacavir insensitive HLA-B*57:03 are shown in magenta for carbon, blue for nitrogen, red for oxygen in panels **C** and **E**. Panel **D** shows that abacavir binding in the F pocket does not alter the peptide conformation compared to other peptide/HLA-B complexes. A cartoon representation of peptide in the crystal structure complexed to abacavir and HLA-B*57:01 is shown in cyan (HSITYLLPV, PDB ID 3UPR). A 9-mer self peptide (LSSPVTKSF) complexed to HLA-B*57:01 (PDB ID: 2RFX, (2)), is shown in red, the 8-mer peptide epitope HIV1 Nef 75-82 (VPLRPMTY) bound to HLA-B*35:01 (PDB ID 1A1N, (40)) is shown in pink, a 9-mer EBV peptide (FLRGRAYGL) complexed to HLA-B8 (PDB ID 1MI5 (41)) is shown in green, and the 11-mer EBV peptide HPVGEADYFEY complexed to HLA-B*35:01 (PDB ID 3MV9, (42)) is shown in yellow. The molecular surface of HLA-B*57:01 from 3UPR is shown in gray. The F pocket residues(9) are colored green, and the A pocket yellow. Panel **E** shows experimental electron density corresponding to abacavir in a Fo-Fc difference map contoured at 3.5σ (red mesh) following molecular replacement. Grey mesh depicts the final 2Fo-Fc electron density map of abacavir in the antigen-binding cleft of HLA-B*57:01 (contour level 1.5σ). H bond interactions between abacavir and HLA-B*57:01 are shown as yellow dashed lines.

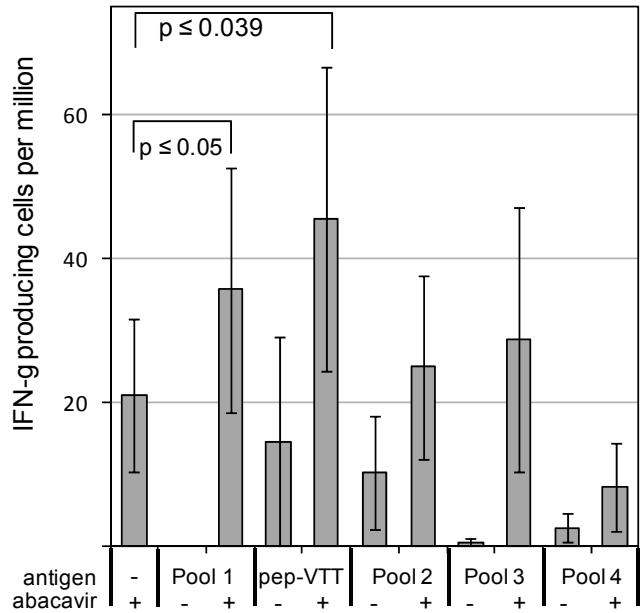

**Figure 4: T cells from hypersensitive donors respond to specific self-peptides in an abacavir dependent fashion.**

PBMCs from five HLA-B*57:01 positive donors with a clinical history of abacavir hypersensitivity were pulsed for 15 minutes with peptide antigens in the presence or absence of abacavir, or with abacavir alone, washed and then tested by IFN-γ ELISpot assay (see methods for details). The figure shows the calculated mean (± SEM) IFN-γ spots per million input PBMC. Statistically significant responses compared to the response induced by the abacavir pulse alone were obtained for peptide pool 1 and the individual peptide VTTDIQVKV (paired two tailed Student t-test on the square root of the SFC counts). Incubating PBMCs from these donors with 10 µg/ml abacavir overnight gave an average of 200 ± 63 SFC per million.

# Supporting Information

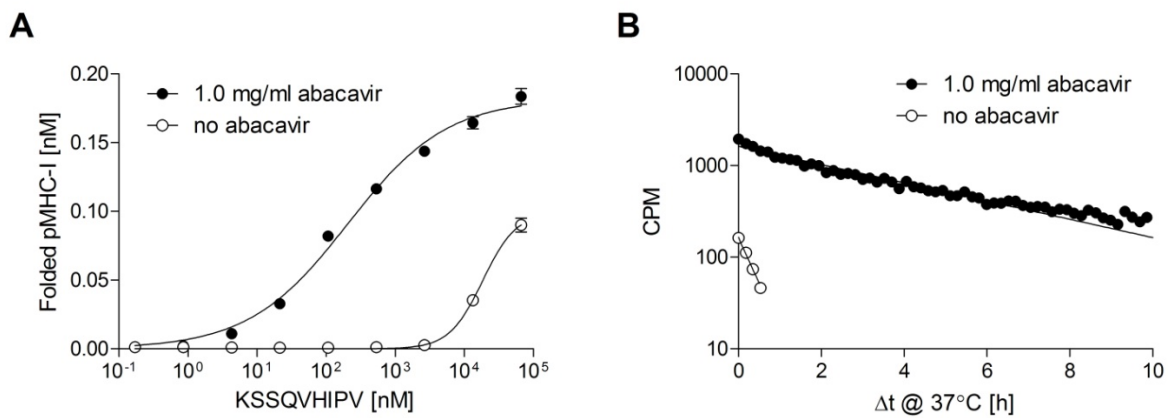

**Figure S1: Binding of pep-V to B*57:01 in AlphaScreen and dissociation assay.**

**A)** The affinity of pep-V (KSSQVHIPV) to biotinylated recombinant B*57:01 was determined with and without adding 1.0 mg/ml abacavir using an AlphaScreen receptor binding assay (1) based upon streptavidin coated donor-beads and W6/32 (anti-HLA class I antibody) coated acceptor-beads. In the presence of abacavir peptide KSSQVHIPV was a good binding peptide, $K_D$=207 nM, whereas in the absence of abacavir peptide KSSQVHIPV was a very weak binder, $K_D$=18,000 nM. **B)** Bound HLA-B*57:01: pep-V complexes are detected in a scintillation proximity assay (SPA, (2)) using radiolabelled B2m and recombinant biotinylated MHC in a streptavidin coated SPA microplate. The signal is generated when B2m* is bound to the bottom of the plate which only occurs in presence of a bound peptide. The dissociation measurement is started by adding excess of unlabelled B2m and raising the temperature to 37C. Measurements are made in 17 minute intervals, creating a dissociation curve. **Table S3** compares different dissociation curves of combinatorial peptide libraries using the area under the dissociation curve (AUC) as a metric divided by the AUC value of the completely randomized library and normalized to an average value of 1.

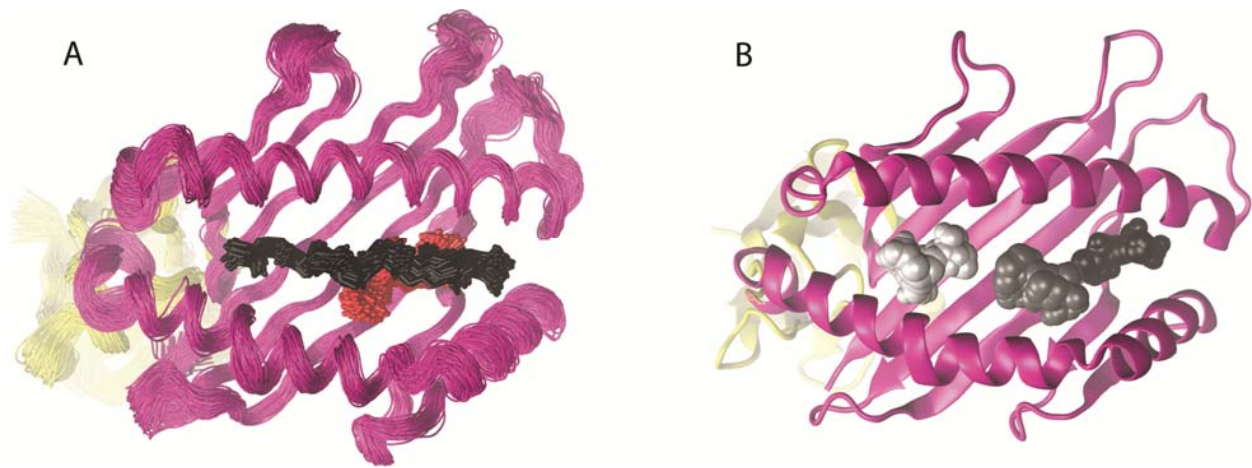

**Figure S2: Computational identification of a binding site for abacavir in HLA-B*57:01.** Panel **A** depicts three potential binding sites for abacavir identified by computational solvent mapping (gray spheres). Panel **B** depicts an ensemble of 100 equally spaced configurations of the complex taken from 30ns of molecular dynamics simulation. Simulations indicate a stable complex on the timescale of the simulation with abacavir localized to the F-pocket, consistent with the subsequent crystal structure (Figure 3).

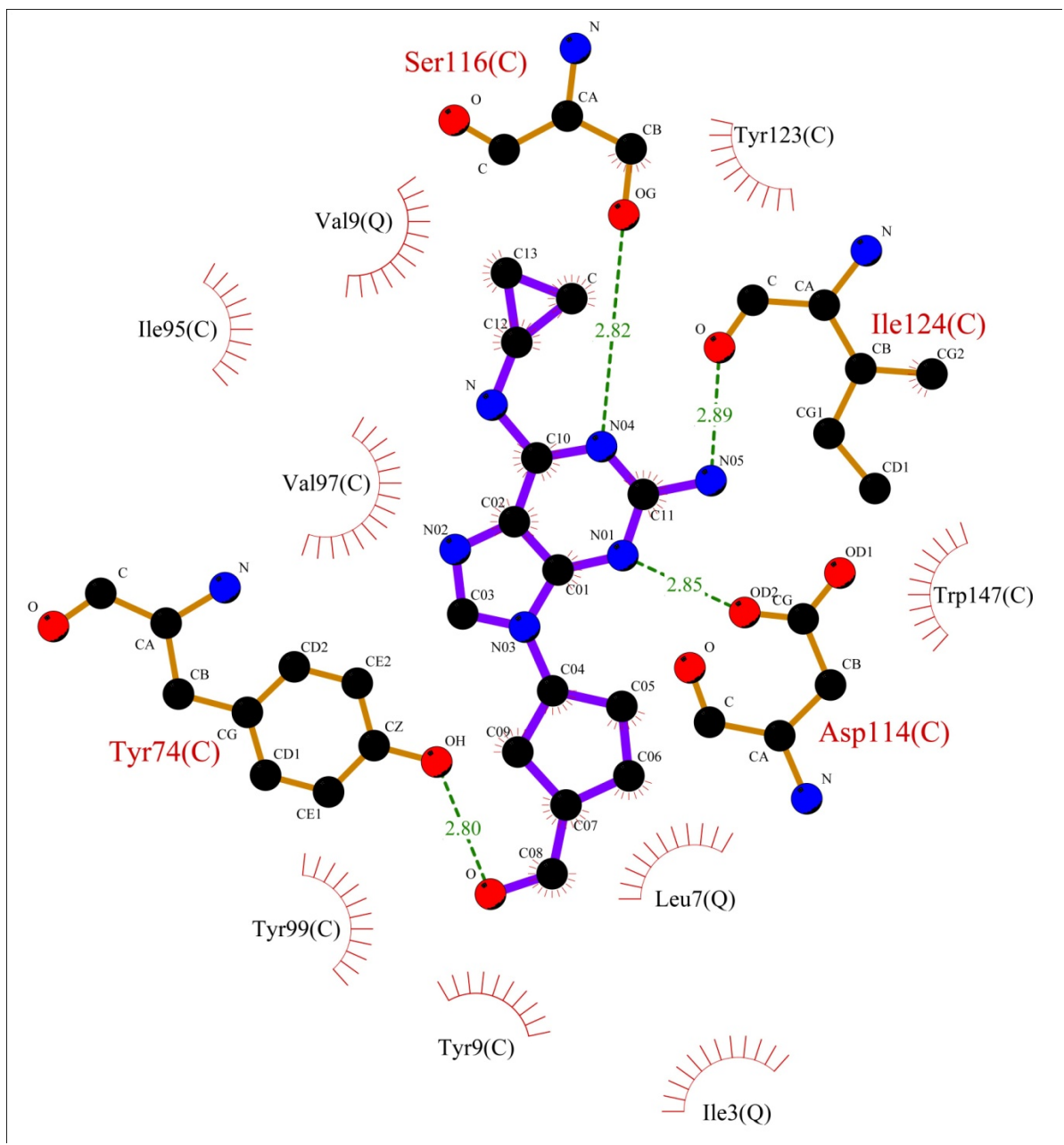

**Figure S3: Intermolecular contacts between abacavir, peptide and HLA-B*57:01.**

Interactions of abacavir, shown in purple, with the surrounding MHC (C chain) and peptide (Q chain) residues based on PDB 3UPR. H bonds are shown as dashed lines and van der Waals contacts are shown as rays. Interactions were calculated with HBplus and Ligplot.

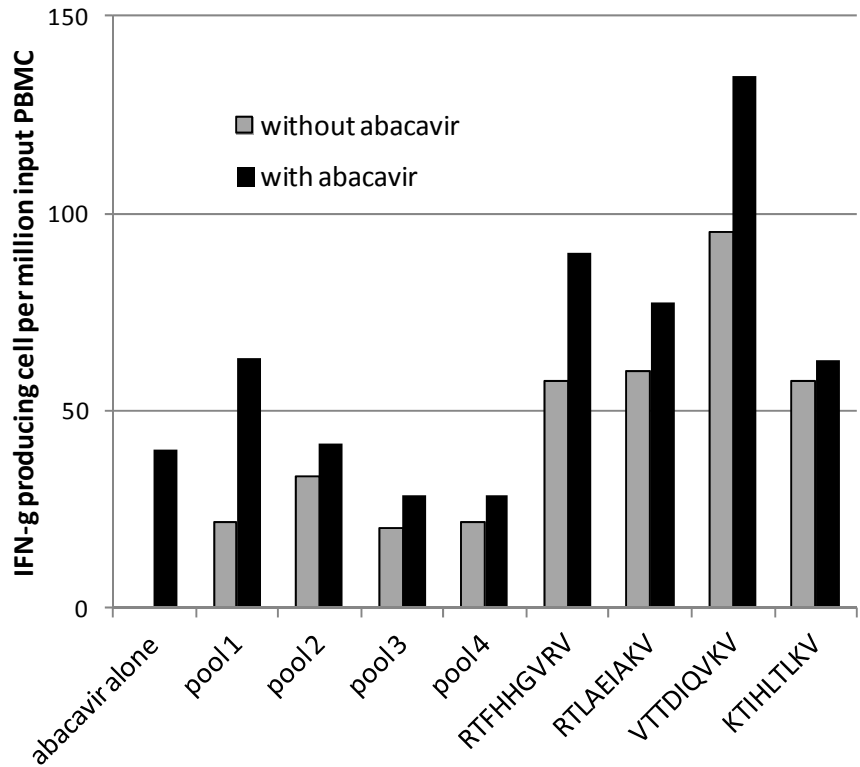

**Figure S4: T cells from hypersensitive donors respond to specific self-peptides in an abacavir dependent fashion.**

As described in Figure 4, PBMCs from a HLA-B*57:01 positive donors with a clinical history of abacavir hypersensitivity were pulsed for 15 minutes with peptide antigens in the presence or absence of abacavir, or with abacavir alone, washed and then tested by IFN-γ ELISpot assay. Responses from one of two donors are shown for whom there was a sufficient amount of PBMCs available to test all four pools of peptides, and also test individual peptides from pool 1 which gave the highest response of all pools. These data were used to select peptide VTTDIQVKV as an individual peptide in all donors (Figure 4).

# Supporting Information References

**Table S1 - Combinatorial library scan for HLA-B*57:01**

| | \multicolumn{9}{c}{Affinity without abcavir [IC50 nM]} | | | | | | | | |
|---|---|---|---|---|---|---|---|---|---|
| | 1 | 2 | 3 | 4 | 5 | 6 | 7 | 8 | 9 |
| A | 11,000 | 2,760 | 982 | 5,830 | 4,470 | 1,030 | 361 | 979 | 11,500 |
| C | 15,400 | 5,220 | 2,660 | 2,960 | 4,770 | 769 | 1,190 | 1,180 | 18,000 |
| D | 15,700 | 1,020 | 9,200 | 3,130 | 5,980 | 1,100 | 2,620 | 5,540 | * |
| E | * | 11,300 | 7,840 | 1,370 | 2,920 | 489 | 833 | 2,010 | * |
| F | * | 7,130 | 1,300 | 1,390 | 889 | 300 | 314 | 1,270 | 287 |
| G | 17,700 | 2,580 | 2,420 | 2,270 | 4,000 | 1,200 | 948 | 1,460 | 11,800 |
| H | 3,620 | 10,600 | 4,370 | 6,040 | 5,660 | 187 | 374 | 921 | 15,500 |
| I | 4,610 | 3,280 | 762 | 1,350 | 3,120 | 814 | 295 | 1,130 | 472 |
| K | 3,610 | 6,640 | 2,390 | 2,480 | 3,420 | 2,390 | 1,460 | 307 | 13,900 |
| L | 9,320 | 1,820 | 821 | 2,420 | 2,010 | 525 | 616 | 739 | 911 |
| M | 6,730 | 1,030 | 685 | 745 | 5,050 | 627 | 754 | 1,300 | 1,190 |
| N | 11,400 | 11,300 | 4,200 | 832 | 8,310 | 608 | 1,020 | 585 | 15,200 |
| P | * | 7,040 | 10,400 | 615 | 3,190 | 591 | 842 | 210 | * |
| Q | 14,500 | 15,200 | 2,530 | 361 | 11,300 | 677 | 345 | 1,250 | * |
| R | 6,470 | 7,270 | 1,400 | 2,080 | 16,300 | 1,270 | 563 | 483 | * |
| S | 8,130 | 147 | 324 | 1,230 | 1,010 | 1,200 | 790 | 673 | * |
| T | 13,400 | 331 | 1,050 | 1,160 | 860 | 917 | 1,590 | 1,620 | * |
| V | * | 469 | 859 | 1,020 | 539 | 582 | 349 | 1,610 | 7,470 |
| W | * | 6,540 | 766 | 1,060 | 691 | 436 | 409 | 1,000 | 176 |
| Y | 13,800 | 4,210 | 2,370 | 1,950 | 925 | 643 | 475 | 668 | 11,800 |

*) >20,000 nM

| | \multicolumn{9}{c}{Affinity with 1mg/ml abcavir [IC50 nM]} | | | | | | | | |
|---|---|---|---|---|---|---|---|---|---|
| | 1 | 2 | 3 | 4 | 5 | 6 | 7 | 8 | 9 |
| A | 12,200 | 2,450 | 541 | 1,560 | 1,030 | 541 | 362 | 678 | 1,720 |
| C | 16,300 | 4,360 | 4,240 | 2,210 | 1,740 | 547 | 703 | 981 | 17,600 |
| D | 15,800 | 1,560 | 3,800 | 2,520 | 2,420 | 562 | 1,340 | 1,790 | * |
| E | 14,700 | 6,680 | 6,810 | 1,560 | 1,900 | 420 | 851 | 934 | 17,800 |
| F | 15,400 | 3,470 | 372 | 1,530 | 715 | 281 | 156 | 590 | 1,370 |
| G | 12,000 | 3,210 | 1,990 | 4,620 | 2,040 | 619 | 939 | 709 | 13,700 |
| H | 3,140 | 5,090 | 2,300 | 1,560 | 3,250 | 317 | 744 | 894 | * |
| I | 7,700 | 960 | 274 | 4,290 | 1,970 | 229 | 142 | 725 | 86 |
| K | 6,760 | 5,660 | 2,840 | 1,540 | 8,140 | 1,210 | 1,910 | 722 | 10,400 |
| L | 10,400 | 1,700 | 666 | 2,110 | 1,140 | 172 | 104 | 624 | 549 |
| M | 9,180 | 1,170 | 454 | 1,220 | 1,470 | 227 | 367 | 842 | 740 |
| N | 12,900 | 6,630 | 2,740 | 1,240 | 7,250 | 271 | 861 | 632 | 17,300 |
| P | 19,500 | 5,890 | 3,210 | 1,370 | 5,540 | 403 | 451 | 142 | * |
| Q | 10,700 | 3,400 | 2,790 | 1,030 | 10,100 | 201 | 545 | 955 | * |
| R | 4,020 | 7,510 | 1,930 | 1,280 | 8,560 | 377 | 963 | 797 | * |
| S | 8,980 | 206 | 478 | 1,020 | 565 | 642 | 994 | 209 | 14,500 |
| T | 9,970 | 369 | 452 | 817 | 601 | 564 | 849 | 1,420 | 17,600 |
| V | 14,100 | 1,130 | 684 | 1,760 | 445 | 395 | 164 | 710 | 851 |
| W | 13,900 | 11,200 | 612 | 697 | 228 | 401 | 160 | 1,060 | 475 |
| Y | 7,590 | 8,330 | 1,070 | 926 | 402 | 554 | 372 | 432 | 16,000 |

*) >20,000 nM

|   | Affinity improvement with abacavir [fold change] | | | | | | | | |
|---|---|---|---|---|---|---|---|---|---|
|   | 1 | 2 | 3 | 4 | 5 | 6 | 7 | 8 | 9 |
| A | - | 1.1 | 1.8 | 3.7 | 4.3 | 1.9 | - | 1.4 | **6.7** |
| C | - | 1.2 | - | 1.3 | 2.7 | 1.4 | 1.7 | 1.2 | 1.0 |
| D | - | - | 2.4 | 1.2 | 2.5 | 1.9 | 2.0 | 3.1 | - |
| E | >1.4 | 1.7 | 1.2 | - | 1.5 | 1.2 | - | 2.1 | >1.1 |
| F | >1.3 | 2.1 | 3.5 | - | 1.2 | 1.1 | 2.0 | 2.2 | - |
| G | 1.5 | - | 1.2 | - | 2.0 | 1.9 | 1.0 | 2.1 | - |
| H | 1.2 | 2.1 | 1.9 | 3.9 | 1.7 | - | - | 1.0 | - |
| I | - | 3.4 | 2.8 | - | 1.6 | 3.5 | 2.1 | 1.6 | **5.5** |
| K | - | 1.2 | - | 1.6 | - | 2.0 | - | - | 1.3 |
| L | - | 1.1 | 1.2 | 1.1 | 1.8 | 3.0 | **5.9** | 1.2 | 1.7 |
| M | - | - | 1.5 | - | 3.4 | 2.8 | 2.1 | 1.5 | 1.6 |
| N | - | 1.7 | 1.5 | - | 1.1 | 2.2 | 1.2 | - | - |
| P | >1 | 1.2 | 3.3 | - | - | 1.5 | 1.9 | 1.5 | - |
| Q | 1.4 | 4.5 | - | - | 1.1 | 3.4 | - | 1.3 | - |
| R | 1.6 | - | - | 1.6 | 1.9 | 3.4 | - | - | - |
| S | - | - | - | 1.2 | 1.8 | 1.9 | - | 3.2 | >1.4 |
| T | 1.3 | - | 2.3 | 1.4 | 1.4 | 1.6 | 1.9 | 1.1 | >1.1 |
| V | >1.4 | - | 1.3 | - | 1.2 | 1.5 | 2.1 | 2.3 | **8.8** |
| W | >1.4 | - | 1.3 | 1.5 | 3.0 | 1.1 | 2.6 | - | - |
| Y | 1.8 | - | 2.2 | 2.1 | 2.3 | 1.2 | 1.3 | 1.5 | - |

**Table S2 - Combinatorial library scan for HLA-B*58:01**

|   | Affinity without abcavir [IC50 nM] | | | | | | | | |
|---|---|---|---|---|---|---|---|---|---|
|   | 1 | 2 | 3 | 4 | 5 | 6 | 7 | 8 | 9 |
| A | * | 748 | 1,290 | 5,260 | 2,920 | 1,320 | 956 | 1,420 | 17,500 |
| C | * | 2,060 | 6,640 | 4,270 | 6,670 | 1,100 | 1,830 | 3,160 | * |
| D | * | 3,320 | 3,050 | 3,360 | 1,450 | 849 | 2,860 | 6,370 | * |
| E | * | 6,340 | 7,540 | 3,900 | 5,500 | 1,890 | 1,150 | 5,790 | * |
| F | * | 1,800 | 1,250 | 2,020 | 2,280 | 696 | 405 | 1,540 | 601 |
| G | * | 2,960 | 1,650 | 8,000 | 1,160 | 1,650 | 1,970 | 1,410 | * |
| H | 11,300 | 9,920 | 7,180 | 8,140 | 3,310 | 1,380 | 1,090 | 1,050 | * |
| I | 17,600 | 3,910 | 1,300 | 6,710 | 4,500 | 755 | 670 | 1,200 | 590 |
| K | 9,710 | 2,450 | 11,800 | 4,030 | 12,900 | 4,260 | 5,980 | 831 | * |
| L | 19,000 | 1,200 | 1,440 | 2,100 | 6,530 | 984 | 1,070 | 865 | 2,430 |
| M | 16,400 | 807 | 1,150 | 1,560 | 3,210 | 760 | 1,120 | 1,260 | 1,280 |
| N | * | 5,510 | 1,580 | 2,710 | 1,280 | 967 | 1,170 | 769 | 13,300 |
| P | * | 2,690 | 13,100 | 1,010 | 4,990 | 952 | 1,240 | 235 | * |
| Q | * | 1,460 | 2,990 | 1,880 | 12,200 | 729 | 675 | 2,770 | * |
| R | 7,410 | 7,790 | 8,240 | 3,910 | 10,000 | 1,560 | 2,030 | 1,340 | * |
| S | 16,300 | 409 | 751 | 2,540 | 1,530 | 1,020 | 1,250 | 580 | * |
| T | * | 1,190 | 1,120 | 1,450 | 1,280 | 1,170 | 2,070 | 5,200 | * |
| V | * | 1,480 | 747 | 1,100 | 1,590 | 916 | 623 | 2,040 | 12,800 |
| W | * | 6,780 | 1,030 | 1,790 | 717 | 1,240 | 525 | 1,790 | 387 |
| Y | * | 7,520 | 4,560 | 2,460 | 426 | 1,430 | 938 | 648 | 11,900 |

*) >20,000 nM

| | \multicolumn{9}{c}{Affinity with 1mg/ml abcavir [IC50 nM]} |
|---|---|---|---|---|---|---|---|---|---|
| | 1 | 2 | 3 | 4 | 5 | 6 | 7 | 8 | 9 |
| A | * | 1,350 | 1,770 | 9,800 | 2,850 | 1,270 | 1,830 | 1,420 | 16,800 |
| C | * | 17,300 | 12,300 | 6,820 | 5,140 | 1,900 | 2,560 | 2,980 | 13,200 |
| D | 17,500 | 16,300 | 5,360 | 7,090 | 7,410 | 1,390 | 6,300 | 9,880 | * |
| E | 19,000 | 9,980 | 7,290 | 6,770 | 2,030 | 977 | 1,270 | 3,610 | * |
| F | * | 9,620 | 1,470 | 3,110 | 5,020 | 811 | 418 | 2,340 | 536 |
| G | * | 6,880 | 2,730 | 5,080 | 6,620 | 5,900 | 4,310 | 1,800 | 17,200 |
| H | 12,400 | 14,300 | 15,300 | 10,800 | 13,400 | 1,490 | 1,290 | 4,790 | * |
| I | 10,200 | 6,300 | 1,870 | 4,710 | 14,500 | 1,640 | 930 | 1,000 | 2,120 |
| K | 16,100 | 15,700 | 14,300 | 9,120 | 4,470 | 9,420 | 5,870 | 1,810 | * |
| L | 13,400 | 8,930 | 2,340 | 4,360 | 7,810 | 793 | 687 | 867 | 2,720 |
| M | 5,600 | 1,880 | 1,400 | 3,130 | 7,950 | 1,100 | 1,200 | 2,470 | 976 |
| N | * | 10,400 | 3,000 | 1,360 | 5,790 | 986 | 966 | 1,240 | 19,700 |
| P | * | 7,150 | 8,760 | 1,650 | 6,290 | 1,660 | 1,130 | 467 | * |
| Q | * | 7,100 | 6,080 | 2,140 | 11,800 | 835 | 1,210 | 3,680 | * |
| R | 12,400 | 4,240 | 10,900 | 6,510 | 15,000 | 3,030 | 3,330 | 2,770 | * |
| S | 13,500 | 421 | 1,190 | 1,620 | 1,540 | 1,190 | 1,690 | 1,140 | * |
| T | 19,200 | 1,750 | 1,310 | 1,550 | 1,340 | 1,180 | 3,880 | 6,920 | * |
| V | * | 2,380 | 1,630 | 2,610 | 2,190 | 725 | 1,120 | 4,120 | * |
| W | * | 10,700 | 1,450 | 2,300 | 1,000 | 1,120 | 754 | 3,960 | 594 |
| Y | 19,900 | 11,500 | 4,680 | 1,720 | 1,490 | 1,670 | 426 | 1,530 | 8,580 |

*) >20,000 nM

| | \multicolumn{9}{c}{Affinity improvement with abacavir [fold change]} |
|---|---|---|---|---|---|---|---|---|---|
| | 1 | 2 | 3 | 4 | 5 | 6 | 7 | 8 | 9 |
| A | - | - | - | - | 1.0 | 1.0 | - | - | 1.0 |
| C | - | - | - | - | 1.3 | - | - | 1.1 | >1.5 |
| D | >1.1 | - | - | - | - | - | - | - | - |
| E | >1.1 | - | 1.0 | - | 2.7 | 1.9 | - | 1.6 | - |
| F | - | - | - | - | - | - | - | - | 1.1 |
| G | - | - | - | 1.6 | - | - | - | - | >1.2 |
| H | - | - | - | - | - | - | - | - | - |
| I | 1.7 | - | - | 1.4 | - | - | - | 1.2 | - |
| K | - | - | - | - | 2.9 | - | 1.0 | - | - |
| L | 1.4 | - | - | - | - | 1.2 | 1.6 | - | - |
| M | 2.9 | - | - | - | - | - | - | - | 1.3 |
| N | - | - | - | 2.0 | - | - | 1.2 | - | - |
| P | - | - | 1.5 | - | - | - | 1.1 | - | - |
| Q | - | - | - | - | 1.0 | - | - | - | - |
| R | - | 1.8 | - | - | - | - | - | - | - |
| S | 1.2 | - | - | 1.6 | - | - | - | - | - |
| T | >1 | - | - | - | - | - | - | - | - |
| V | - | - | - | - | - | 1.3 | - | - | - |
| W | - | - | - | - | - | 1.1 | - | - | - |
| Y | >1 | - | - | 1.4 | - | - | 2.2 | - | 1.4 |

**Table S3: C-terminal combinatorial library scan of HLA B*57:01 and HLA B*57:03 in the dissociation assay.**

| Peptide | HLA B*57:01 | | | HLA B*57:03 | | |
|---|---|---|---|---|---|---|
| | -abacavir | +abacavir | Diff^2 | -abacavir | +abacavir | Diff^2 |
| X8A | 0.0 | 1.0 | **0.8** | 0.2 | 0.2 | 0.0 |
| X8C | 0.3 | 0.7 | 0.1 | 1.0 | 1.0 | 0.0 |
| X8D | 0.1 | 0.0 | 0.0 | 0.1 | 0.1 | 0.0 |
| X8E | 0.2 | 0.0 | 0.0 | 0.1 | 0.1 | 0.0 |
| X8F | 4.2 | 1.8 | **5.8** | 4.4 | 4.2 | 0.0 |
| X8G | 0.2 | 0.4 | 0.1 | 0.1 | 0.1 | 0.0 |
| X8H | 0.5 | 0.0 | 0.2 | 0.3 | 0.2 | 0.0 |
| X8I | 1.5 | 4.2 | **7.2** | 2.2 | 2.1 | 0.0 |
| X8K | 0.1 | 0.0 | 0.0 | 0.1 | 0.1 | 0.0 |
| X8L | 0.7 | 1.6 | **0.9** | 1.8 | 1.7 | 0.0 |
| X8M | 1.2 | 2.2 | **1.0** | 2.1 | 2.0 | 0.0 |
| X8N | 0.0 | 0.1 | 0.0 | 0.1 | 0.0 | 0.0 |
| X8P | 0.0 | 0.4 | 0.2 | 0.2 | 0.2 | 0.0 |
| X8Q | 0.0 | 0.1 | 0.0 | 0.1 | 0.1 | 0.0 |
| X8R | 0.1 | 0.1 | 0.0 | 0.1 | 0.1 | 0.0 |
| X8S | 0.4 | 0.3 | 0.0 | 0.1 | 0.1 | 0.0 |
| X8T | 0.2 | 0.5 | 0.1 | 0.1 | 0.1 | 0.0 |
| X8V | 0.3 | 2.0 | **2.8** | 0.9 | 1.0 | 0.0 |
| X8W | 7.7 | 3.3 | **18.8** | 4.8 | 4.9 | 0.0 |
| X8Y | 2.2 | 1.1 | **1.4** | 1.3 | 1.6 | 0.1 |

Dissociation curves were generated as explained in Supplemental Figure 3. The reported values are area under the dissociation curve (AUC) divided by the AUC value of the completely randomized library and normalized to an average value of 1 per column

## Table S4: X-ray data collection and statistics

| Field | Value |
|---|---|
| **Protein** | HLA B*57:01•β-microglobulin•pep-V•abacavir |
| **PDB Accession Code** | 3UPR |
| **Space Group** | P 1 2$_1$ 1 |
| **Unit Cell Dimensions** | |
|   a, b, c (Å) | 44.75, 131.41, 88.34 |
|   α,β,γ (°) | 90, 104.30, 90 |
| **Resolution (Å)** | 30.00 – 2.00 |
| **Number of unique Reflections** | 64,702 (2605) |
| **Completeness (%)** | 96.85 (73.00) |
| **Redundancy** | 4.30 (2.70) |
| **R$_{merge}$** [1] | 0.090 (0.48) |
| **I/σ(I)** | 22.84 (2.2) |
| **R$_{work}$** [2], **R$_{free}$** [3] | 0.18, 0.22 |
| **Ramachandran Statistics** [4] **(%)** | |
|   Most Favored | 96.82 |
|   Allowed | 3.18 |
|   Outliers | 0 |
| **Number of atoms** | |
|   Protein, solvent | 6287, 431 |
|   Abacavir | 42 |
| **Average B factors (Å$^2$)** | |
|   Protein (heavy chain, light chain) | 35.45, 33.00 |
|   Waters | 33.48 |
|   Pep-V | 27.51 |
|   Abacavir | 21.61 |

Values in parentheses are for highest resolution bin.

[1] Rmerge = Σ |I - <I>|/ Σ <I>.

[2] R$_{work}$ = Σ|F$_o$(hkl)| - |F$_c$(hkl)|/ Σ |F$_o$(hkl)|.

[3] R$_{free}$ is calculated in the same manner based on 10% of the reflection data not used during refinement.

[4] Statistics generated using MOLPROBITY (*Chen et al. (2010) Acta Cryst D66:12-21* ).

**Table S5: List of peptides eluted in the presence and absence of abacavir.**

| Sample | Peptide Sequence* | C-terminus | Length | Accession | Source Protein | Start | Stop |
|---|---|---|---|---|---|---|---|
| both | KTKRSIQF | F | 8 | P68366 | Tubulin alpha-4A chain | 336 | 343 |
| both | RTRIGYSF | F | 8 | Q9NYH9 | U3 small nucleolar RNA-associated protein 6 homolog | 72 | 79 |
| both | GTHKVTVLF | F | 9 | P21333 | Filamin-A | 352 | 360 |
| both | HTQKNRQFF | F | 9 | O95766 | UPF0550 protein C7orf28 | 85 | 93 |
| both | ISKKINTKF | F | 9 | O00232 | 26S proteasome non-ATPase regulatory subunit 12 | 205 | 213 |
| both | ISNKKEAKF | F | 9 | Q9UBU7 | Protein DBF4 homolog A | 87 | 95 |
| both | ISTPVIRTF | F | 9 | Q9NZB2 | Constitutive coactivator of PPAR-gamma-like protein 1 | 989 | 997 |
| both | ITNKYQLVF | F | 9 | Q9NYV6 | RNA polymerase I-specific transcription initiation factor RRN3 | 516 | 524 |
| both | KAKKEIEKF | F | 9 | Q9UHW5 | GPN-loop GTPase 3 | 182 | 190 |
| both | KARQKTNVF | F | 9 | Q9ULX3 | RNA-binding protein NOB1 | 354 | 362 |
| both | KSKLSQNNF | F | 9 | Q9Y277 | Voltage-dependent anion-selective channel protein 3 | 161 | 169 |
| both | KSKLTRNNF | F | 9 | P45880 | Voltage-dependent anion-selective channel protein 2 | 172 | 180 |
| both | KSLGVTTKF | F | 9 | Q5VWZ2 | Lysophospholipase-like protein 1 | 195 | 203 |
| both | KSRELNTRF | F | 9 | Q8IXQ4 | Uncharacterized protein KIAA1704 | 323 | 331 |
| both | KSRGVLHQF | F | 9 | P27144 | Adenylate kinase isoenzyme 4, mitochondrial | 186 | 194 |
| both | KSRVTQSNF | F | 9 | P21796 | Voltage-dependent anion-selective channel protein 1 | 161 | 169 |
| both | KSSDIAKTF | F | 9 | P13797 | Plastin-3 | 82 | 90 |
| both | KSSQVQRRF | F | 9 | P13010 | ATP-dependent DNA helicase 2 subunit 2 | 347 | 355 |
| both | KSSRKTYTF | F | 9 | P52732 | Kinesin-like protein KIF11 | 60 | 68 |
| both | KSTDVAKTF | F | 9 | P13796 | Plastin-2 | 82 | 90 |
| both | KTAPFDSRF | F | 9 | P14854 | Cytochrome c oxidase subunit VIb isoform 1 | 13 | 21 |
| both | KTKNAVEKF | F | 9 | Q7Z7A1 | Centriolin | 1455 | 1463 |
| both | KTKQGGLRF | F | 9 | Q96SZ6 | CDK5 regulatory subunit-associated protein 1 | 346 | 354 |
| both | KTKRDINRF | F | 9 | Q9HAS0 | Protein Njmu-R1 | 238 | 246 |
| both | KTRTNVPTF | F | 9 | Q96BU1 | S100P-binding protein | 305 | 313 |
| both | KTSGIINKF | F | 9 | O95243 | Methyl-CpG-binding domain protein 4 | 328 | 336 |
| both | LSSPVTKSF | F | 9 | P01834 | Ig kappa chain C region | 93 | 101 |
| both | RAKPEYHKF | F | 9 | Q00341 | Vigilin | 733 | 741 |
| both | RARTIYERF | F | 9 | Q9BZJ0 | Crooked neck-like protein 1 | 361 | 369 |
| both | RSLRLQKQF | F | 9 | Q12788 | WD repeat-containing protein SAZD | 208 | 216 |
| both | RSSGVRRSF | F | 9 | Q14667 | UPF0378 protein KIAA0100 | 2083 | 2091 |
| both | RTHPVATSF | F | 9 | Q96B26 | Exosome complex exonuclease RRP43 | 190 | 198 |
| both | RTKENAEKF | F | 9 | Q14894 | Mu-crystallin homolog | 169 | 177 |
| both | RTKRSSVVF | F | 9 | Q92608 | Dedicator of cytokinesis protein 2 | 1701 | 1709 |
| both | RTQIRHMKF | F | 9 | O95503 | Chromobox protein homolog 6 | 221 | 229 |
| both | RTRKGILLF | F | 9 | Q04941 | Proteolipid protein 2 | 21 | 29 |
| both | RTSIVEKRF | F | 9 | Q9BXR0 | Queuine tRNA-ribosyltransferase | 345 | 353 |
| both | RTVEKPPKF | F | 9 | O15020 | Spectrin beta chain, brain 2 | 351 | 359 |
| both | TTLEHQKTF | F | 9 | Q13347 | Eukaryotic translation initiation factor 3 subunit I | 218 | 226 |
| both | VSKAPAPTF | F | 9 | A6NCL7 | Ankyrin repeat domain-containing protein LOC651746 | 428 | 436 |
| both | VSKPSAPVF | F | 9 | Q69YN4 | Protein virilizer homolog | 76 | 84 |
| both | VSKRLHQRF | F | 9 | Q96ST3 | Paired amphipathic helix protein Sin3a | 1202 | 1210 |
| both | VTHSVRIGF | F | 9 | P26010 | Integrin beta-7 | 185 | 193 |
| both | YTYRGPKAF | F | 9 | Q12874 | Splicing factor 3A subunit 3 | 414 | 422 |
| both | KAHPPELKKF | F | 10 | P62308 | Small nuclear ribonucleoprotein G | 3 | 12 |
| both | KARREAKVKF | F | 10 | P61353 | 60S ribosomal protein L27 | 109 | 118 |
| both | KSRGYVKEQF | F | 10 | P46783 | 40S ribosomal protein S10 | 53 | 62 |
| both | KTKDGVREVF | F | 10 | P61586 | Transforming protein RhoA | 162 | 171 |
| both | KTKEGVREVF | F | 10 | P08134 | Rho-related GTP-binding protein RhoC | 162 | 171 |
| both | KTKNAQQVHF | F | 10 | Q66K89 | Transcription factor E4F1 | 529 | 538 |
| both | KTLDQAIMKF | F | 10 | Q92878 | DNA repair protein RAD50 | 1126 | 1135 |
| both | KTLHDTRTHF | F | 10 | Q8IZT6 | Abnormal spindle-like microcephaly-associated protein | 1862 | 1871 |
| both | KTNEHGMKTF | F | 10 | Q562F6 | Shugoshin-like 2 | 389 | 398 |
| both | KTRRPDNTAF | F | 10 | Q9NV96 | Cell cycle control protein 50A | 24 | 33 |
| both | KTVEPTGKRF | F | 10 | P10155 | 60 kDa SS-A/Ro ribonucleoprotein | 362 | 371 |
| both | LSKQETGTEF | F | 10 | Q15051 | IQ calmodulin-binding motif-containing protein 1 | 261 | 270 |
| both | LSRPDGSASF | F | 10 | Q9NQT4 | Exosome complex exonuclease RRP46 | 38 | 47 |
| both | STAKEIIQHF | F | 10 | Q9UET6 | Putative ribosomal RNA methyltransferase 1 | 96 | 105 |
| both | VTKTVSNDSF | F | 10 | P55209 | Nucleosome assembly protein 1-like 1 | 288 | 297 |
| both | ITSSIHSKETF | F | 11 | Q07864 | DNA polymerase epsilon catalytic subunit A | 1890 | 1900 |
| both | KSKEDLVSQGF | F | 11 | Q96FW1 | Ubiquitin thioesterase OTUB1 | 120 | 130 |
| both | KSKPVEKNYAF | F | 11 | P09884 | DNA polymerase alpha catalytic subunit | 431 | 441 |

| Sample | Peptide Sequence* | C-terminus | Length | Accession | Source Protein | Start | Stop |
|---|---|---|---|---|---|---|---|
| both | KTKSTETEHSF | F | 11 | Q9UHC1 | DNA mismatch repair protein Mlh3 | 617 | 627 |
| both | KTLVNPANVTF | F | 11 | O75608 | Acyl-protein thioesterase 1 | 190 | 200 |
| both | KTMSAKEKGKF | F | 11 | B2RPK0 | Putative high mobility group protein 1-like 1 | 50 | 60 |
| both | KTMSAKEKSKF | F | 11 | P26583 | 60S ribosomal protein L29 | 50 | 60 |
| both | RTISTASRRHF | F | 11 | P14406 | Cytochrome c oxidase polypeptide 7A2, mitochondrial | 14 | 24 |
| both | RTNPNSGDFRF | F | 11 | Q9NVW2 | E3 ubiquitin-protein ligase RNF12 | 127 | 137 |
| both | HSIQYSGEVRRF | F | 12 | P04040 | Catalase | 421 | 432 |
| both | KARELDPRI | I | 9 | Q9ULW0 | Targeting protein for Xklp2 | 403 | 411 |
| both | KSAIPHPLI | I | 9 | P61201 | COP9 signalosome complex subunit 2 | 225 | 233 |
| both | KSKNILFVI | I | 9 | O15069 | NAC-alpha domain-containing protein 1 | 1439 | 1447 |
| both | KSKPDKQLI | I | 9 | Q03933 | Heat shock factor protein 2 | 435 | 443 |
| both | KSRDYEREI | I | 9 | Q9NRZ9 | Lymphoid-specific helicase | 776 | 784 |
| both | KSVLVKQTI | I | 9 | Q9UHX1 | Poly | 80 | 88 |
| both | KTAVVVGTI | I | 9 | Q07020 | 60S ribosomal protein L18 | 78 | 86 |
| both | KTKGFHTTI | I | 9 | Q15417 | Calponin-3 | 130 | 138 |
| both | KTKKSLESI | I | 9 | P62888 | 60S ribosomal protein L30 | 6 | 14 |
| both | KTKPIHTII | I | 9 | Q8IWI9 | MAX gene-associated protein | 1032 | 1040 |
| both | KVRTEKETI | I | 9 | P12270 | Nucleoprotein TPR | 1405 | 1413 |
| both | LSKRNPRQI | I | 9 | P83731 | 60S ribosomal protein L24 | 41 | 49 |
| both | KARLPLRL | L | 8 | O00767 | Acyl-CoA desaturase | 129 | 136 |
| both | AAPEGKRSL | L | 9 | Q9Y6B2 | EP300-interacting inhibitor of differentiation 1 | 67 | 75 |
| both | HTWNGIRHL | L | 9 | Q99643 | Succinate dehydrogenase cytochrome b560 subunit, mitochondrial | 127 | 135 |
| both | ISKQFHHQL | L | 9 | Q99613 | Eukaryotic translation initiation factor 3 subunit C | 710 | 718 |
| both | KTFRIKRFL | L | 9 | P62891 | 60S ribosomal protein L39 | 5 | 13 |
| both | KTIKLWNTL | L | 9 | P63244 | Guanine nucleotide-binding protein subunit beta-2-like 1 | 127 | 135 |
| both | KTRMEERRL | L | 9 | Q9BY42 | UPF0549 protein C20orf43 | 183 | 191 |
| both | RAREYNARL | L | 9 | Q00653 | Nuclear factor NF-kappa-B p100 subunit | 442 | 450 |
| both | RSNRVVRTL | L | 9 | Q9Y5Y5 | Peroxisomal membrane protein PEX16 | 170 | 178 |
| both | RTKASVHTL | L | 9 | O43660 | Pleiotropic regulator 1 | 318 | 326 |
| both | TTSRVLKVL | L | 9 | P35222 | Catenin beta-1 | 339 | 347 |
| both | KTWAGSQSRRL | L | 11 | Q9UI43 | Putative ribosomal RNA methyltransferase 2 | 194 | 204 |
| both | RTMSSSDRAMM | M | 11 | Q00403 | Transcription initiation factor IIB | 105 | 115 |
| both | KTKDVTVW | W | 8 | Q96DR4 | StAR-related lipid transfer protein 4 | 34 | 41 |
| both | LSKKSQKW | W | 8 | P41236 | Protein phosphatase inhibitor 2 | 40 | 47 |
| both | MAARGHSW | W | 8 | Q14213 | Interleukin-27 subunit beta | 70 | 77 |
| both | VSKVSTTW | W | 8 | Q9UKT4 | F-box only protein 5 | 274 | 281 |
| both | VTRKSWLW | W | 8 | P82930 | 28S ribosomal protein S34, mitochondrial | 80 | 87 |
| both | VTRQGALW | W | 8 | Q5SRD1 | Putative mitochondrial import inner membrane translocase subunit | 120 | 127 |
| both | AAQITQRKW | W | 9 | P30504 | HLA class I histocompatibility antigen, Cw-4 alpha chain | 163 | 171 |
| both | ASHKGQKLW | W | 9 | O15049 | NEDD4-binding protein 3 | 135 | 143 |
| both | ETKKKVFLW | W | 9 | A6NDK8 | Protein FAM35B | 480 | 488 |
| both | FVKKLEHSW | W | 9 | Q99755 | Phosphatidylinositol-4-phosphate 5-kinase type-1 alpha | 358 | 366 |
| both | GSKDKNFQW | W | 9 | P49792 | E3 SUMO-protein ligase RanBP2 | 2864 | 2872 |
| both | GTDKTLRTW | W | 9 | Q9NWT1 | p21-activated protein kinase-interacting protein 1 | 142 | 150 |
| both | GTLKGHNGW | W | 9 | P63244 | Guanine nucleotide-binding protein subunit beta-2-like 1 | 9 | 17 |
| both | GTLSGHASW | W | 9 | Q9GZS3 | WD repeat-containing protein 61 | 226 | 234 |
| both | GTVLKTSSW | W | 9 | Q460N5 | Poly [ADP-ribose] polymerase 14 | 993 | 1001 |
| both | GTYGVRAAW | W | 9 | P13598 | Intercellular adhesion molecule 2 | 258 | 266 |
| both | HSHPHITVW | W | 9 | P46736 | BRCA1/BRCA2-containing complex subunit 3 | 122 | 130 |
| both | IAAPDSRRW | W | 9 | O95297 | Myelin protein zero-like protein 1 | 11 | 19 |
| both | IAAQTGTRW | W | 9 | Q9BZL1 | Ubiquitin-like protein 5 | 31 | 39 |
| both | ISKEEAMRW | W | 9 | P62913 | 60S ribosomal protein L11 | 157 | 165 |
| both | ISSSAGARW | W | 9 | Q92673 | Sortilin-related receptor | 522 | 530 |
| both | ITAGAHRLW | W | 9 | O00767 | Acyl-CoA desaturase | 115 | 123 |
| both | ITKKQVSVW | W | 9 | P13196 | 5-aminolevulinate synthase, nonspecific, mitochondrial | 240 | 248 |
| both | ITTKAISRW | W | 9 | P49368 | T-complex protein 1 subunit gamma | 160 | 168 |
| both | ITYDKLNKW | W | 9 | P05120 | Plasminogen activator inhibitor 2 | 286 | 294 |
| both | ITYQHIDRW | W | 9 | Q9UBQ5 | Eukaryotic translation initiation factor 3 subunit K | 148 | 156 |
| both | KAAPRSQHW | W | 9 | Q92917 | G patch domain and KOW motifs-containing protein | 349 | 357 |
| both | KAKREVSSW | W | 9 | P53582 | Methionine aminopeptidase 1 | 59 | 67 |
| both | KATPIKLIW | W | 9 | P49685 | G-protein coupled receptor 15 | 187 | 195 |
| both | KGKPIRIMW | W | 9 | Q13310 | Polyadenylate-binding protein 4 | 78 | 86 |
| both | KGKPVRIMW | W | 9 | P11940 | Polyadenylate-binding protein 1 | 78 | 86 |

| Sample | Peptide Sequence* | C-terminus | Length | Accession | Source Protein | Start | Stop |
|---|---|---|---|---|---|---|---|
| both | KLKDIRNAW | W | 9 | Q9UKD2 | mRNA turnover protein 4 homolog | 52 | 60 |
| both | KSFENISKW | W | 9 | P61026 | Ras-related protein Rab-10 | 95 | 103 |
| both | KSGSVQEQW | W | 9 | Q9NXF1 | Testis-expressed sequence 10 protein | 895 | 903 |
| both | KSIRNDIEW | W | 9 | Q9NPA3 | Mid1-interacting protein 1 | 95 | 103 |
| both | KSKPKNSVW | W | 9 | Q92963 | GTP-binding protein Rit1 | 196 | 204 |
| both | KSSLTQHSW | W | 9 | Q9NYB0 | Telomeric repeat-binding factor 2-interacting protein 1 | 169 | 177 |
| both | KTHPLWRLW | W | 9 | Q9HBE5 | Interleukin-21 receptor | 256 | 264 |
| both | KTKETSVNW | W | 9 | P55265 | Double-stranded RNA-specific adenosine deaminase | 1120 | 1128 |
| both | KTKEVIQEW | W | 9 | Q9Y490 | Talin-1 | 343 | 351 |
| both | KTKEVLQEW | W | 9 | Q9Y4G6 | Talin-2 | 346 | 354 |
| both | KVKEQKDYW | W | 9 | Q13137 | Calcium-binding and coiled-coil domain-containing protein 2 | 197 | 205 |
| both | LAAVRGEQW | W | 9 | P13498 | Cytochrome b-245 light chain | 123 | 131 |
| both | LATTILQHW | W | 9 | P78527 | DNA-dependent protein kinase catalytic subunit | 1618 | 1626 |
| both | LSGPFVQKW | W | 9 | P22102 | Trifunctional purine biosynthetic protein adenosine-3 | 899 | 907 |
| both | LSQEQLRQW | W | 9 | P32248 | C-C chemokine receptor type 7 | 347 | 355 |
| both | LTYENVERW | W | 9 | Q15907 | Ras-related protein Rab-11B | 97 | 105 |
| both | LVRNSHHTW | W | 9 | Q9NXG6 | Putative HIF-prolyl hydroxylase PH-4 | 271 | 279 |
| both | RADKRVNEW | W | 9 | Q15393 | Splicing factor 3B subunit 3 | 528 | 536 |
| both | RAKNVRNDW | W | 9 | Q6RFH5 | WD repeat-containing protein 74 | 169 | 177 |
| both | RGKAHAAVW | W | 9 | O00170 | AH receptor-interacting protein | 271 | 279 |
| both | RSKEESAHW | W | 9 | Q9BY42 | UPF0549 protein C20orf43 | 290 | 298 |
| both | RSRsPHRKW | W | 9 | Q9NYV4 | Cell division cycle 2-related protein kinase 7 | 219 | 227 |
| both | RSRsRERHW | W | 9 | Q16560 | U11/U12 small nuclear ribonucleoprotein 35 kDa protein | 186 | 194 |
| both | RTVIIEQSW | W | 9 | P10809 | 60 kDa heat shock protein, mitochondrial | 60 | 68 |
| both | RVAQMKRTW | W | 9 | Q8IY37 | Probable ATP-dependent RNA helicase DHX37 | 835 | 843 |
| both | RVHPETYEW | W | 9 | Q7KZ85 | Transcription elongation factor SPT6 | 1057 | 1065 |
| both | SADKTVKLW | W | 9 | Q9Y263 | Phospholipase A-2-activating protein | 170 | 178 |
| both | SADRTVKLW | W | 9 | P57737 | Coronin-7 | 97 | 105 |
| both | SSEKHAYSW | W | 9 | Q8WUM0 | Nuclear pore complex protein Nup133 | 302 | 310 |
| both | SSLKKKQIW | W | 9 | Q16658 | Fascin | 38 | 46 |
| both | SSSKEHVKW | W | 9 | P46063 | ATP-dependent DNA helicase Q1 | 164 | 172 |
| both | SSSRIRAAW | W | 9 | P14543 | Nidogen-1 | 4 | 12 |
| both | STKRKAAVW | W | 9 | P21359 | Neurofibromin | 259 | 267 |
| both | SVKRNQAVW | W | 9 | Q6ZNB6 | NF-X1-type zinc finger protein NFXL1 | 166 | 174 |
| both | SVTEIQEKW | W | 9 | Q9NR30 | Nucleolar RNA helicase 2 | 689 | 697 |
| both | TAKISDFSW | W | 9 | Q16576 | Histone-binding protein RBBP7 | 373 | 381 |
| both | TSKITSTAW | W | 9 | Q9UKA1 | F-box/LRR-repeat protein 5 | 423 | 431 |
| both | TTFSNSQRW | W | 9 | O14966 | Ras-related protein Rab-7L1 | 94 | 102 |
| both | TTSKPVRKW | W | 9 | Q8IWZ6 | Bardet-Biedl syndrome 7 protein | 222 | 230 |
| both | VAHTIKQAW | W | 9 | Q95460 | Major histocompatibility complex class I-related gene protein | 157 | 165 |
| both | VARKSEEHW | W | 9 | Q8TAT5 | Endonuclease VIII-like 3 | 313 | 321 |
| both | VSKTLPSTW | W | 9 | Q14677 | Clathrin interactor 1 | 469 | 477 |
| both | VTALAARTW | W | 9 | Q9UII2 | ATPase inhibitor, mitochondrial | 3 | 11 |
| both | VTKKTYEIW | W | 9 | Q14839 | Chromodomain-helicase-DNA-binding protein 4 | 1721 | 1729 |
| both | VTNKSVSVW | W | 9 | Q9H974 | Queuine tRNA-ribosyltransferase domain-containing protein 1 | 108 | 116 |
| both | VTNPHSSQW | W | 9 | P24522 | Growth arrest and DNA-damage-inducible protein GADD45 alpha | 127 | 135 |
| both | VTYKNVPNW | W | 9 | P62826 | GTP-binding nuclear protein Ran | 96 | 104 |
| both | YTDNLVRVW | W | 9 | P63244 | Guanine nucleotide-binding protein subunit beta-2-like 1 | 302 | 310 |
| both | ASDDQTIRVW | W | 10 | P53621 | Coatomer subunit alpha | 111 | 120 |
| both | ASKGREAMEW | W | 10 | Q9Y3B1 | Protein slowmo homolog 2 | 153 | 162 |
| both | ATRKHGTDLW | W | 10 | Q9NP61 | ADP-ribosylation factor GTPase-activating protein 3 | 124 | 133 |
| both | AVTALAARTW | W | 10 | Q9UII2 | ATPase inhibitor, mitochondrial | 2 | 11 |
| both | ETKKDHPYTW | W | 10 | Q9Y570 | Protein phosphatase methylesterase 1 | 278 | 287 |
| both | GTDRNHNRYW | W | 10 | Q9UIG0 | Bromodomain adjacent to zinc finger domain protein 1B | 904 | 913 |
| both | HGKPVTQVTW | W | 10 | Q14137 | Ribosome biogenesis protein BOP1 | 533 | 542 |
| both | HSAHTKLQTW | W | 10 | Q96MD2 | UPF0536 protein C12orf66 | 217 | 226 |
| both | HSITSDNHKW | W | 10 | Q8NEM7 | Protein FAM48A | 178 | 187 |
| both | HTAPPENKTW | W | 10 | Q6N021 | Protein TET2 | 994 | 1003 |
| both | HTLDDRTQLW | W | 10 | Q9BXP5 | Arsenite-resistance protein 2 | 529 | 538 |
| both | IAANEENRKW | W | 10 | O75368 | SH3 domain-binding glutamic acid-rich-like protein | 39 | 48 |
| both | ITAENVAKKW | W | 10 | Q9BWD1 | Acetyl-CoA acetyltransferase, cytosolic | 162 | 171 |
| both | ITKADAAEFW | W | 10 | Q13191 | E3 ubiquitin-protein ligase CBL-B | 173 | 182 |
| both | ITKEIEANEW | W | 10 | O75410 | Transforming acidic coiled-coil-containing protein 1 | 615 | 624 |

| Sample | Peptide Sequence* | C-terminus | Length | Accession | Source Protein | Start | Stop |
|---|---|---|---|---|---|---|---|
| both | ITSQDVLHSW | W | 10 | P00403 | Cytochrome c oxidase subunit 2 | 154 | 163 |
| both | KAHQVKNHMW | W | 10 | P11388 | DNA topoisomerase 2-alpha | 352 | 361 |
| both | KANEGKKETW | W | 10 | Q96K76 | Ubiquitin carboxyl-terminal hydrolase 47 | 997 | 1006 |
| both | KAREYSKEGW | W | 10 | Q5XKP0 | Protein QIL1 | 101 | 110 |
| both | KARVETQNHW | W | 10 | P28066 | Proteasome subunit alpha type-5 | 91 | 100 |
| both | KAVIDLNNRW | W | 10 | Q01081 | Splicing factor U2AF 35 kDa subunit | 125 | 134 |
| both | KGFSEEHNTW | W | 10 | P45973 | Chromobox protein homolog 5 | 42 | 51 |
| both | KSAEKEISLW | W | 10 | P22392 | Nucleoside diphosphate kinase B | 124 | 133 |
| both | KSGPVVSLGW | W | 10 | Q9H269 | Vacuolar protein sorting-associated protein 16 homolog | 82 | 91 |
| both | KSKGSQTEEW | W | 10 | Q14202 | Zinc finger MYM-type protein 3 | 851 | 860 |
| both | KSSIHDARSW | W | 10 | Q8N2I9 | Serine/threonine-protein kinase 40 | 389 | 398 |
| both | KTEASETRKW | W | 10 | Q1ED39 | Protein C16orf88 | 345 | 354 |
| both | KTNIQKEATW | W | 10 | A8MYJ9 | Importin subunit alpha-2-like protein | 109 | 118 |
| both | KTQTPNRKEW | W | 10 | P09874 | Poly [ADP-ribose] polymerase 1 | 324 | 333 |
| both | KVAKEAANRW | W | 10 | Q9BWT6 | Meiotic nuclear division protein 1 homolog | 160 | 169 |
| both | MTAGVDGHSW | W | 10 | O75694 | Nuclear pore complex protein Nup155 | 450 | 459 |
| both | RAKTDSAEKW | W | 10 | Q15058 | Kinesin-like protein KIF14 | 124 | 133 |
| both | RALELEARRW | W | 10 | Q03169 | Tumor necrosis factor, alpha-induced protein 2 | 334 | 343 |
| both | RATQSGQRRW | W | 10 | Q5SXM2 | snRNA-activating protein complex subunit 4 | 722 | 731 |
| both | RQNPHHVHEW | W | 10 | Q9HCS7 | Pre-mRNA-splicing factor SYF1 | 351 | 360 |
| both | RSFKETTNRW | W | 10 | Q92613 | Protein Jade-3 | 714 | 723 |
| both | RSQKAEGQRW | W | 10 | O15533 | Tapasin | 339 | 348 |
| both | RSRDISREEW | W | 10 | Q9ULV3 | Cip1-interacting zinc finger protein | 767 | 776 |
| both | RSYSESEKQW | W | 10 | Q96HQ2 | CDKN2AIP N-terminal-like protein | 28 | 37 |
| both | RTAGHPLTRW | W | 10 | P35236 | Tyrosine-protein phosphatase non-receptor type 7 | 76 | 85 |
| both | RTILVDNNTW | W | 10 | P46977 | Dolichyl-diphosphooligosaccharide--protein glycosyltransferase | 538 | 547 |
| both | RTTLVDNNTW | W | 10 | Q8TCJ2 | Dolichyl-diphosphooligosaccharide--protein glycosyltransferase | 617 | 626 |
| both | RVKPLHYISW | W | 10 | O43847 | Nardilysin | 453 | 462 |
| both | RVKSVNLDQW | W | 10 | Q8IYB5 | Stromal membrane-associated protein 1 | 69 | 78 |
| both | SAASPHYQEW | W | 10 | Q6SPF0 | Atherin | 25 | 34 |
| both | SATDAAIRVW | W | 10 | Q12788 | WD repeat-containing protein SAZD | 57 | 66 |
| both | TAAQITQRKW | W | 10 | P30480 | HLA class I histocompatibility antigen, B-42 alpha chain | 162 | 171 |
| both | TSSKPDPSQW | W | 10 | O75410 | Transforming acidic coiled-coil-containing protein 1 | 380 | 389 |
| both | TVYTGIDHHW | W | 10 | Q9NV06 | WD repeat and SOF domain-containing protein 1 | 157 | 166 |
| both | VTNKSQIRTW | W | 10 | P27694 | Replication protein A 70 kDa DNA-binding subunit | 203 | 212 |
| both | VTVPANVQRW | W | 10 | Q13155 | Multisynthetase complex auxiliary component p38 | 293 | 302 |
| both | AAFPAEKESEW | W | 11 | Q9BRF8 | Uncharacterized metallophosphoesterase CSTP1 | 19 | 29 |
| both | ASYSGKAADVW | W | 11 | Q96RU7 | Tribbles homolog 3 | 235 | 245 |
| both | GTWIGKGTERW | W | 11 | Q9C0C9 | Ubiquitin-conjugating enzyme E2 O | 1045 | 1055 |
| both | GTYMGHTGAVW | W | 11 | Q13347 | Eukaryotic translation initiation factor 3 subunit I | 46 | 56 |
| both | HSNEQTLQRSW | W | 11 | Q9Y6J8 | Serine/threonine/tyrosine-interacting-like protein 1 | 263 | 273 |
| both | HTNQDHVHAVW | W | 11 | Q9UNY4 | Transcription termination factor 2 | 526 | 536 |
| both | KASSEVERQRW | W | 11 | P22059 | Oxysterol-binding protein 1 | 162 | 172 |
| both | KIRsPGKNHKW | W | 11 | Q9UHK0 | Nuclear fragile X mental retardation-interacting protein 1 | 289 | 299 |
| both | KSKAMTGVEQW | W | 11 | P49368 | T-complex protein 1 subunit gamma | 425 | 435 |
| both | KSKLDAEVSKW | W | 11 | P35221 | Catenin alpha-1 | 695 | 705 |
| both | KTVDGPSGKLW | W | 11 | P04406 | Glyceraldehyde-3-phosphate dehydrogenase | 186 | 196 |
| both | KVARDQAVKKW | W | 11 | O75792 | Ribonuclease H2 subunit A | 183 | 193 |
| both | KVIHEQVNHRW | W | 11 | P11388 | DNA topoisomerase 2-alpha | 287 | 297 |
| both | QTDPSGTYHAW | W | 11 | Q8TAA3 | Proteasome subunit alpha type-7-like | 154 | 164 |
| both | RALEAEKRALW | W | 11 | Q14160 | Protein LAP4 | 1491 | 1501 |
| both | RARLEESRRLW | W | 11 | Q01082 | Spectrin beta chain, brain 1 | 631 | 641 |
| both | RTIQTPIGSTW | W | 11 | Q9BVJ6 | U3 small nucleolar RNA-associated protein 14 homolog A | 697 | 707 |
| both | RVFEDESGKHW | W | 11 | Q8TEA8 | D-tyrosyl-tRNA | 53 | 63 |
| both | RVRAGPKKESW | W | 11 | Q96GQ5 | UPF0420 protein C16orf58 | 407 | 417 |
| both | SASPDATIRIW | W | 11 | Q9UMS4 | Pre-mRNA-processing factor 19 | 281 | 291 |
| both | SSATDAAIRVW | W | 11 | Q12788 | WD repeat-containing protein SAZD | 56 | 66 |
| both | TGSTDKTVRLW | W | 11 | O75529 | TAF5-like RNA polymerase II p300/CBP-associated factor | 443 | 453 |
| both | TSRPPAQGASW | W | 11 | O14976 | Cyclin G-associated kinase | 1121 | 1131 |
| both | VAKKTKDVTVW | W | 11 | Q96DR4 | StAR-related lipid transfer protein 4 | 31 | 41 |
| both | VGLPAAGKTTW | W | 11 | Q9BUJ2 | Heterogeneous nuclear ribonucleoprotein U-like protein 1 | 427 | 437 |
| both | VVSPHEDMRTW | W | 11 | P42345 | FKBP12-rapamycin complex-associated protein | 1643 | 1653 |
| both | RSKDDPGKGSYW | W | 12 | Q9UPW0 | Forkhead box protein J3 | 141 | 152 |

| Sample | Peptide Sequence* | C-terminus | Length | Accession | Source Protein | Start | Stop |
|---|---|---|---|---|---|---|---|
| both | RTKANEGKKETW | W | 12 | Q96K76 | Ubiquitin carboxyl-terminal hydrolase 47 | 995 | 1006 |
| both | RTLDEAVGVQKW | W | 12 | Q14667 | UPF0378 protein KIAA0100 | 836 | 847 |
| both | VSDSGAHVLNSW | W | 12 | Q9NVM9 | Uncharacterized protein C12orf11 | 80 | 91 |
| both | KVKDGPGGKEATW | W | 13 | P22307 | Non-specific lipid-transfer protein | 462 | 474 |
| both | KARNSFRY | Y | 8 | P46779 | 60S ribosomal protein L28 | 33 | 40 |
| both | HTRSDVRLY | Y | 9 | Q86TI2 | Dipeptidyl peptidase 9 | 614 | 622 |
| both | ISKALVAYY | Y | 9 | P62249 | 40S ribosomal protein S16 | 88 | 96 |
| both | KAAEMRRIY | Y | 9 | Q9NUP1 | Protein cappuccino homolog | 124 | 132 |
| both | KASGTLREY | Y | 9 | Q02543 | 60S ribosomal protein L18a | 2 | 10 |
| both | KSFVKVYNY | Y | 9 | P61353 | 60S ribosomal protein L27 | 69 | 77 |
| both | KSKITHPVY | Y | 9 | Q9UGP8 | Translocation protein SEC63 homolog | 631 | 639 |
| both | KSRRDTPKY | Y | 9 | Q9Y2X3 | Nucleolar protein 5 | 330 | 338 |
| both | KTAGKRYVY | Y | 9 | P14921 | Protein C-ets-1 | 404 | 412 |
| both | KTFPYQHRY | Y | 9 | Q8NI27 | THO complex subunit 2 | 398 | 406 |
| both | KTKEIEQVY | Y | 9 | Q8NEY8 | Periphilin-1 | 297 | 305 |
| both | KTKFPAEQY | Y | 9 | Q15631 | Translin | 76 | 84 |
| both | KTNQLHRMY | Y | 9 | Q9UBI1 | COMM domain-containing protein 3 | 137 | 145 |
| both | KTRIIDVVY | Y | 9 | P62241 | 40S ribosomal protein S8 | 75 | 83 |
| both | LAKTGVHHY | Y | 9 | P62888 | 60S ribosomal protein L30 | 66 | 74 |
| both | LVKREDYLY | Y | 9 | O95298 | NADH dehydrogenase [ubiquinone] 1 subunit C2 | 74 | 82 |
| both | MTNPHNHLY | Y | 9 | Q15006 | Tetratricopeptide repeat protein 35 | 183 | 191 |
| both | RSQVQRARY | Y | 9 | P60900 | Proteasome subunit alpha type-6 | 88 | 96 |
| both | RSRRDELHY | Y | 9 | Q9Y5W7 | Sorting nexin-14 | 184 | 192 |
| both | RTIVIRRDY | Y | 9 | P62280 | 40S ribosomal protein S11 | 84 | 92 |
| both | RVKQEEQRY | Y | 9 | O75410 | Transforming acidic coiled-coil-containing protein 1 | 722 | 730 |
| both | RVRDVVTKY | Y | 9 | Q12931 | Heat shock protein 75 kDa, mitochondrial | 269 | 277 |
| both | HTFQNDIHVY | Y | 10 | Q15392 | 24-dehydrocholesterol reductase | 399 | 408 |
| both | KIYPGHGRRY | Y | 10 | P83731 | 60S ribosomal protein L24 | 12 | 21 |
| both | KSFSESGINY | Y | 10 | O60762 | Dolichol-phosphate mannosyltransferase | 49 | 58 |
| both | KTAVIDHHNY | Y | 10 | P39656 | Dolichyl-diphosphooligosaccharide--protein glycosyltransferase | 136 | 145 |
| both | KTKEVIGTGY | Y | 10 | Q9BT09 | Protein canopy homolog 3 | 68 | 77 |
| both | KTLPADVQNY | Y | 10 | Q99590 | SFRS2-interacting protein | 837 | 846 |
| both | KSKPDKQLIQY | Y | 11 | Q03933 | Heat shock factor protein 2 | 435 | 445 |
| both | KTLPETSLPNY | Y | 11 | Q9BYI3 | Hyccin | 17 | 27 |
| both | RTKKVGIVGKY | Y | 11 | P61513 | Outer dense fiber protein 2 | 4 | 14 |
| both | DGSSLPADVHRY | Y | 12 | Q96KA5 | Cleft lip and palate transmembrane protein 1-like protein | 186 | 197 |
| treated | RSVALAVLA | A | 9 | P61769 | Beta-2-microglobulin | 3 | 11 |
| treated | ATHPSRAQF | F | 9 | O95834 | Echinoderm microtubule-associated protein-like 2 | 384 | 392 |
| treated | HSNPEPKTF | F | 9 | Q9NS87 | Kinesin-like protein KIF15 | 62 | 70 |
| treated | KSKENPRNF | F | 9 | P37802 | Transgelin-2 | 154 | 162 |
| treated | KSVKKGFEF | F | 9 | Q15019 | Septin-2 | 30 | 38 |
| treated | KSYGNEKRF | F | 9 | Q06330 | Recombining binding protein suppressor of hairless | 58 | 66 |
| treated | KVAEVTKKF | F | 9 | P35914 | Hydroxymethylglutaryl-CoA lyase, mitochondrial | 184 | 192 |
| treated | RSKTVYEGF | F | 9 | Q9NW38 | E3 ubiquitin-protein ligase FANCL | 20 | 28 |
| treated | RSQQNRHSF | F | 9 | O76039 | Cyclin-dependent kinase-like 5 | 439 | 447 |
| treated | RSYEEKKQF | F | 9 | Q6IQ32 | ADNP homeobox protein 2 | 1049 | 1057 |
| treated | RTATFQQRF | F | 9 | Q9BY77 | Polymerase delta-interacting protein 3 | 46 | 54 |
| treated | RTHTGEKKF | F | 9 | Q13118 | Krueppel-like factor 10 | 421 | 429 |
| treated | RTHTGEKRF | F | 9 | Q8TD94 | Krueppel-like factor 14 | 247 | 255 |
| treated | RTKDILIRF | F | 9 | Q68CP4 | Heparan-alpha-glucosaminide N-acetyltransferase | 527 | 535 |
| treated | RTKQVLHTF | F | 9 | Q9BTV5 | Fibronectin type III and SPRY domain-containing protein 1 | 439 | 447 |
| treated | RVAQKKASF | F | 9 | P46777 | 60S ribosomal protein L5 | 279 | 287 |
| treated | SSRRKQLTF | F | 9 | Q8WTV1 | THAP domain-containing protein 3 | 15 | 23 |
| treated | STTDVEKSF | F | 9 | Q53EL6 | Programmed cell death protein 4 | 232 | 240 |
| treated | VAKKIHEEF | F | 9 | Q9Y2L1 | Exosome complex exonuclease RRP44 | 674 | 682 |
| treated | VSKSVTQGF | F | 9 | Q92599 | Septin-8 | 35 | 43 |
| treated | HSGDHGKRLF | F | 10 | O43674 | NADH dehydrogenase] 1 beta subcomplex subunit 5, mitochondrial | 46 | 55 |
| treated | KSSSHERRAF | F | 10 | P51532 | Probable global transcription activator SNF2L4 | 1237 | 1246 |
| treated | KTKSENGLEF | F | 10 | P21796 | Voltage-dependent anion-selective channel protein 1 | 32 | 41 |
| treated | KTMKMRGQAF | F | 10 | P08579 | U2 small nuclear ribonucleoprotein B'' | 44 | 53 |
| treated | KTRSTKDQQF | F | 10 | O75044 | SLIT-ROBO Rho GTPase-activating protein 2 | 77 | 86 |
| treated | RSDTSDKRKF | F | 10 | P20591 | Interferon-induced GTP-binding protein Mx1 | 633 | 642 |
| treated | RTKQQREAEF | F | 10 | Q15005 | Signal peptidase complex subunit 2 | 180 | 189 |

| Sample | Peptide Sequence* | C-terminus | Length | Accession | Source Protein | Start | Stop |
|---|---|---|---|---|---|---|---|
| treated | KSRKEQKETTF | F | 11 | P35408 | Prostaglandin E2 receptor EP4 subtype | 44 | 54 |
| treated | KSRPEDQRSSF | F | 11 | Q9UPZ3 | Hermansky-Pudlak syndrome 5 protein | 899 | 909 |
| treated | KSRSSRAGLQF | F | 11 | Q8IUE6 | Histone H2A type 2-B | 16 | 26 |
| treated | KTRSKGRNEQF | F | 11 | O75691 | Small subunit processome component 20 homolog | 484 | 494 |
| treated | RSRPQKHQNTF | F | 11 | Q96MD7 | Uncharacterized protein C9orf85 | 10 | 20 |
| treated | KSYKGRQI | I | 8 | Q14683 | Structural maintenance of chromosomes protein 1A | 13 | 20 |
| treated | RTKKPVVI | I | 8 | Q9NSE4 | Isoleucyl-tRNA synthetase, mitochondrial | 468 | 475 |
| treated | ASVGDRVTI | I | 9 | P04432 | Ig kappa chain V-I region Daudi | 35 | 43 |
| treated | ATIDKVVKI | I | 9 | Q13535 | Serine/threonine-protein kinase ATR | 561 | 569 |
| treated | AVITNQPEI | I | 9 | P25963 | NF-kappa-B inhibitor alpha | 118 | 126 |
| treated | GTHVGKVNI | I | 9 | Q9GZS3 | WD repeat-containing protein 61 | 124 | 132 |
| treated | ISIKQEPKI | I | 9 | Q96CB8 | Integrator complex subunit 12 | 65 | 73 |
| treated | ITKTVVENI | I | 9 | P57740 | Nuclear pore complex protein Nup107 | 635 | 643 |
| treated | KAIHNKVNI | I | 9 | Q15019 | Septin-2 | 169 | 177 |
| treated | KAYSEAHEI | I | 9 | P61981 | 14-3-3 protein gamma | 152 | 160 |
| treated | KSATKQPSI | I | 9 | O94782 | Ubiquitin carboxyl-terminal hydrolase 1 | 343 | 351 |
| treated | KSIKNIQKI | I | 9 | P36542 | ATP synthase subunit gamma, mitochondrial | 36 | 44 |
| treated | KSKSEEIKI | I | 9 | Q96CS2 | Coiled-coil domain-containing protein 5 | 130 | 138 |
| treated | KTFTTQETI | I | 9 | P06744 | Glucose-6-phosphate isomerase | 211 | 219 |
| treated | KTIAEGRRI | I | 9 | P09132 | Signal recognition particle 19 kDa protein | 27 | 35 |
| treated | KTIDGQQTI | I | 9 | P52907 | F-actin-capping protein subunit alpha-1 | 146 | 154 |
| treated | KTKDGVVEI | I | 9 | P04792 | Heat shock protein beta-1 | 112 | 120 |
| treated | KTKVQEEKI | I | 9 | Q8TCG1 | Protein CIP2A | 807 | 815 |
| treated | KTKVQKDEI | I | 9 | Q7Z333 | Probable helicase senataxin | 779 | 787 |
| treated | KTSGSNVKI | I | 9 | P56381 | ATP synthase subunit epsilon, mitochondrial | 37 | 45 |
| treated | MQNPRQYKI | I | 9 | P62269 | 40S ribosomal protein S18 | 71 | 79 |
| treated | NTLPTKETI | I | 9 | P63313 | Thymosin beta-10 | 27 | 35 |
| treated | NTVELRVKI | I | 9 | P04844 | Dolichyl-diphosphooligosaccharide--protein glycosyltransferase | 361 | 369 |
| treated | QVFKGAVKI | I | 9 | P37268 | Squalene synthetase | 308 | 316 |
| treated | RSMQNSPSI | I | 9 | Q8N201 | Integrator complex subunit 1 | 2109 | 2117 |
| treated | RTQHVKVWI | I | 9 | Q9BZJ0 | Crooked neck-like protein 1 | 705 | 713 |
| treated | RTVKDAHSI | I | 9 | Q8NAV1 | Pre-mRNA-splicing factor 38A | 4 | 12 |
| treated | SSAKIVPKI | I | 9 | Q7Z417 | Nuclear fragile X mental retardation-interacting protein 2 | 348 | 356 |
| treated | STTRVKPFI | I | 9 | Q9Y6A4 | UPF0468 protein C16orf80 | 110 | 118 |
| treated | SVAQVKAMI | I | 9 | O15205 | Ubiquitin D | 111 | 119 |
| treated | VGARIYHTI | I | 9 | P10620 | Microsomal glutathione S-transferase 1 | 111 | 119 |
| treated | VSTGLKVRI | I | 9 | P13639 | Elongation factor 2 | 421 | 429 |
| treated | VTGVTRVTI | I | 9 | Q9H009 | Nascent polypeptide-associated complex subunit alpha-2 | 88 | 96 |
| treated | GTASVHKVTI | I | 10 | O60244 | Mediator of RNA polymerase II transcription subunit 14 | 350 | 359 |
| treated | KAERERGITI | I | 10 | P68104 | Elongation factor 1-alpha 1 | 64 | 73 |
| treated | KSAPELKTGI | I | 10 | P19338 | Nucleolin | 318 | 327 |
| treated | KSYEAQDPEI | I | 10 | Q99622 | Protein C10 | 86 | 95 |
| treated | KTYQDIQNTI | I | 10 | P24928 | DNA-directed RNA polymerase II subunit RPB1 | 697 | 706 |
| treated | LTTKLTNTNI | I | 10 | P00403 | Cytochrome c oxidase subunit 2 | 46 | 55 |
| treated | RGVPDAKIRI | I | 10 | P27635 | 60S ribosomal protein L10 | 24 | 33 |
| treated | RVAQLEQVYI | I | 10 | P62318 | Small nuclear ribonucleoprotein Sm D3 | 54 | 63 |
| treated | TTVEDLGSKI | I | 10 | P35613 | Basigin | 144 | 153 |
| treated | KAISNKDQHSI | I | 11 | Q96FA3 | Protein pellino homolog 1 | 67 | 77 |
| treated | KTFEEKQGTEI | I | 11 | P19338 | Nucleolin | 444 | 454 |
| treated | VTYPAKAKGTFI | I | 12 | P04844 | Dolichyl-diphosphooligosaccharide--protein glycosyltransferase | 396 | 407 |
| treated | KSRESYVL | L | 8 | Q14249 | Endonuclease G, mitochondrial | 76 | 83 |
| treated | RSKKAHVL | L | 8 | P35221 | Catenin alpha-1 | 54 | 61 |
| treated | ITAHVEPLL | L | 9 | Q15306 | Interferon regulatory factor 4 | 401 | 409 |
| treated | KAASVRPVL | L | 9 | Q9H269 | Vacuolar protein sorting-associated protein 16 homolog | 58 | 66 |
| treated | KAKPVTTNL | L | 9 | Q8N0T1 | Uncharacterized protein C8orf59 | 28 | 36 |
| treated | KAVSLQRKL | L | 9 | Q14966 | Zinc finger protein 638 | 658 | 666 |
| treated | KSMSGRQKL | L | 9 | O15212 | Prefoldin subunit 6 | 25 | 33 |
| treated | KSRKNNYGL | L | 9 | O76021 | Ribosomal L1 domain-containing protein 1 | 48 | 56 |
| treated | KTFRDKNTL | L | 9 | Q9NRM2 | Zinc finger protein 277 | 219 | 227 |
| treated | KTFVQKQTL | L | 9 | Q8N1W2 | Zinc finger protein 710 | 526 | 534 |
| treated | KTKAAVVTL | L | 9 | Q8IZT6 | Abnormal spindle-like microcephaly-associated protein | 2019 | 2027 |
| treated | KVIEINPYL | L | 9 | P28074 | Proteasome subunit beta type-5 | 92 | 100 |
| treated | LAPERRSTL | L | 9 | Q9NX31 | Uncharacterized protein C20orf111 | 226 | 234 |

| Sample | Peptide Sequence* | C-terminus | Length | Accession | Source Protein | Start | Stop |
|---|---|---|---|---|---|---|---|
| treated | LSKPNPPSL | L | 9 | Q9UKK3 | Poly [ADP-ribose] polymerase 4 | 364 | 372 |
| treated | LTKKSSETL | L | 9 | P09543 | 2',3'-cyclic-nucleotide 3'-phosphodiesterase | 194 | 202 |
| treated | QSAKTQIKL | L | 9 | O60343 | TBC1 domain family member 4 | 444 | 452 |
| treated | QTKRVKANL | L | 9 | P35579 | Myosin-9 | 1207 | 1215 |
| treated | RSRTIVHTL | L | 9 | Q9NWU5 | 39S ribosomal protein L22, mitochondrial | 198 | 206 |
| treated | RSTQVKKQL | L | 9 | O60281 | Zinc finger protein 292 | 1247 | 1255 |
| treated | RTAQKTALL | L | 9 | Q9BVM2 | Protein DPCD | 11 | 19 |
| treated | SAPSRATAL | L | 9 | Q6PI47 | BTB/POZ domain-containing protein KCTD18 | 326 | 334 |
| treated | STIRLLTSL | L | 9 | P49368 | T-complex protein 1 subunit gamma | 458 | 466 |
| treated | STLHLVLRL | L | 9 | P62988 | Ubiquitin | 65 | 73 |
| treated | KTKADGSFRL | L | 10 | Q5JPE7 | Nodal modulator 2 | 358 | 367 |
| treated | VTATDIRVTL | L | 10 | P11047 | Laminin subunit gamma-1 | 244 | 253 |
| treated | KSKDIVNKM | M | 9 | P52732 | Kinesin-like protein KIF11 | 757 | 765 |
| treated | KSKNILVRM | M | 9 | O75394 | 39S ribosomal protein L33, mitochondrial | 12 | 20 |
| treated | QTARQHPKM | M | 9 | P01374 | Lymphotoxin-alpha | 46 | 54 |
| treated | STRKIHTTM | M | 9 | Q9H6R4 | Nucleolar protein 6 | 361 | 369 |
| treated | RAFHHGRFIM | M | 10 | Q14534 | Squalene monooxygenase | 222 | 231 |
| treated | RTAKGEKFVM | M | 10 | P30101 | Protein disulfide-isomerase A3 | 329 | 338 |
| treated | KSHSHtPSRRM | M | 11 | Q9UQ35 | Serine/arginine repetitive matrix protein 2 | 471 | 481 |
| treated | ASANLRVLV | V | 9 | Q8WW24 | Tektin-4 | 262 | 270 |
| treated | ATIKLQSTV | V | 9 | Q8IZT6 | Abnormal spindle-like microcephaly-associated protein | 1681 | 1689 |
| treated | HSIPVTVEV | V | 9 | Q9UQL6 | Histone deacetylase 5 | 26 | 34 |
| treated | KAAKIRVSV | V | 9 | Q6P1K8 | General transcription factor IIH subunit 2-like protein | 190 | 198 |
| treated | KIYEGQVEV | V | 9 | P46777 | 60S ribosomal protein L5 | 117 | 125 |
| treated | KSNGTIIHV | V | 9 | P10747 | T-cell-specific surface glycoprotein CD28 | 127 | 135 |
| treated | KTIHLTLKV | V | 9 | O15205 | Ubiquitin D | 72 | 80 |
| treated | KVAKVEPAV | V | 9 | Q9P2E9 | Ribosome-binding protein 1 | 145 | 153 |
| treated | RSARVTVAV | V | 9 | Q9UDX5 | Mitochondrial fission protein MTP18 | 73 | 81 |
| treated | RTFHHGVRV | V | 9 | Q8NFH4 | Nucleoporin Nup37 | 67 | 75 |
| treated | RTLAEIAKV | V | 9 | Q15233 | Non-POU domain-containing octamer-binding protein | 119 | 127 |
| treated | RTTETQVLV | V | 9 | P12081 | Histidyl-tRNA synthetase, cytoplasmic | 405 | 413 |
| treated | RVAGIHKKV | V | 9 | P26373 | 60S ribosomal protein L13 | 82 | 90 |
| treated | TVAPFNPTV | V | 9 | Q96BU1 | S100P-binding protein | 147 | 155 |
| treated | VTTDIQVKV | V | 9 | O00267 | Transcription elongation factor SPT5 | 980 | 988 |
| treated | ATKRAQAW | W | 8 | P51149 | Ras-related protein Rab-7a | 135 | 142 |
| treated | HSARSHKW | W | 8 | Q6ZNA4 | E3 ubiquitin-protein ligase Arkadia | 177 | 184 |
| treated | KSKNSIQW | W | 8 | Q15329 | Transcription factor E2F5 | 107 | 114 |
| treated | KSKSETRW | W | 8 | Q9Y520 | BAT2 domain-containing protein 1 | 986 | 993 |
| treated | KTTVHRKW | W | 8 | P46063 | ATP-dependent DNA helicase Q1 | 353 | 360 |
| treated | LTNKGHKW | W | 8 | Q9Y222 | Cyclin-D-binding Myb-like transcription factor 1 | 144 | 151 |
| treated | QSRRPSSW | W | 8 | Q13526 | Peptidyl-prolyl cis-trans isomerase NIMA-interacting 1 | 66 | 73 |
| treated | RSKKAVDW | W | 8 | Q8WUM9 | Sodium-dependent phosphate transporter 1 | 637 | 644 |
| treated | STKESPRW | W | 8 | Q9Y561 | Low-density lipoprotein receptor-related protein 12 | 6 | 13 |
| treated | AANHKVAKW | W | 9 | Q9NR09 | Baculoviral IAP repeat-containing protein 6 | 180 | 188 |
| treated | ASKVTEQEW | W | 9 | P09497 | Clathrin light chain B | 120 | 128 |
| treated | ATDAAIRVW | W | 9 | Q12788 | WD repeat-containing protein SAZD | 58 | 66 |
| treated | ATSRGAAGW | W | 9 | Q8WY91 | THAP domain-containing protein 4 | 119 | 127 |
| treated | ATVTIQRHW | W | 9 | Q8IZT6 | Abnormal spindle-like microcephaly-associated protein | 1396 | 1404 |
| treated | GTNAKVRTW | W | 9 | Q9H0H0 | Integrator complex subunit 2 | 308 | 316 |
| treated | HARGSSRSW | W | 9 | Q9HCU9 | Breast cancer metastasis-suppressor 1 | 186 | 194 |
| treated | HATGHHYTW | W | 9 | P22059 | Oxysterol-binding protein 1 | 567 | 575 |
| treated | HSRQGSDEW | W | 9 | Q5PRF9 | Sterile alpha motif domain-containing protein 4B | 167 | 175 |
| treated | HSRsRsPQW | W | 9 | Q9UQ35 | Serine/arginine repetitive matrix protein 2 | 505 | 513 |
| treated | HTSTSKRHW | W | 9 | Q4ZIN3 | Membralin | 396 | 404 |
| treated | ISKRLGKRW | W | 9 | P35716 | Transcription factor SOX-11 | 79 | 87 |
| treated | KAAEKKARW | W | 9 | Q8IV08 | Phospholipase D3 | 30 | 38 |
| treated | KAGPGLKRW | W | 9 | Q96N66 | Membrane-bound O-acyltransferase domain-containing protein 7 | 24 | 32 |
| treated | KAGQVVTIW | W | 9 | P02545 | Lamin-A/C | 490 | 498 |
| treated | KAKVTGDEW | W | 9 | P13807 | Glycogen [starch] synthase, muscle | 50 | 58 |
| treated | KAVARRKNW | W | 9 | O75821 | Eukaryotic translation initiation factor 3 subunit G | 94 | 102 |
| treated | KLQEKARKW | W | 9 | Q6P2Q9 | Pre-mRNA-processing-splicing factor 8 | 29 | 37 |
| treated | KSAKNPERW | W | 9 | Q96P70 | Importin-9 | 900 | 908 |
| treated | KSFDNIRNW | W | 9 | P61006 | Ras-related protein Rab-8A | 94 | 102 |

| Sample | Peptide Sequence* | C-terminus | Length | Accession | Source Protein | Start | Stop |
|---|---|---|---|---|---|---|---|
| treated | KSKKREERW | W | 9 | Q7Z7A1 | Centriolin | 1336 | 1344 |
| treated | KSNPKNDSW | W | 9 | Q96PP9 | Guanylate-binding protein 4 | 121 | 129 |
| treated | KSSGAPASW | W | 9 | Q8IX18 | Probable ATP-dependent RNA helicase DHX40 | 564 | 572 |
| treated | KTFTGHREW | W | 9 | P43034 | Platelet-activating factor acetylhydrolase IB subunit alpha | 228 | 236 |
| treated | KTPsQRNKW | W | 9 | Q8IVF7 | Formin-like protein 3 | 957 | 965 |
| treated | KTSDPGRSW | W | 9 | Q96JI7 | Spatacsin | 405 | 413 |
| treated | LAALRHARW | W | 9 | Q9UPR6 | Zinc finger RNA-binding protein 2 | 769 | 777 |
| treated | MSKEKKEEW | W | 9 | Q08945 | FACT complex subunit SSRP1 | 588 | 596 |
| treated | MTNGFHMTW | W | 9 | Q68CL5 | Tubulin polyglutamylase complex subunit 2 | 71 | 79 |
| treated | NSKRKAETW | W | 9 | Q9Y2H1 | Serine/threonine-protein kinase 38-like | 264 | 272 |
| treated | NSNRDRREW | W | 9 | O95104 | Splicing factor, arginine/serine-rich 15 | 1025 | 1033 |
| treated | QSLKSTAKW | W | 9 | O43665 | Regulator of G-protein signaling 10 | 22 | 30 |
| treated | RSHRHTDSW | W | 9 | Q9H089 | Large subunit GTPase 1 homolog | 25 | 33 |
| treated | RSHRRDQKW | W | 9 | P62266 | 40S ribosomal protein S23 | 14 | 22 |
| treated | RSKAEGKRW | W | 9 | Q9UNY4 | Transcription termination factor 2 | 80 | 88 |
| treated | RSKsRSRSW | W | 9 | O75494 | FUS-interacting serine-arginine-rich protein 1 | 246 | 254 |
| treated | RTKEEVNEW | W | 9 | Q9UKD2 | mRNA turnover protein 4 homolog | 107 | 115 |
| treated | RTKQRAREW | W | 9 | Q8IY17 | Neuropathy target esterase | 1022 | 1030 |
| treated | RTREEGKDW | W | 9 | Q5JSZ5 | Uncharacterized protein KIAA0515 | 427 | 435 |
| treated | SAKKASTLW | W | 9 | Q9P275 | Ubiquitin carboxyl-terminal hydrolase 36 | 692 | 700 |
| treated | SSARGSRLW | W | 9 | Q8WUJ1 | Cytochrome b5 domain-containing protein 2 | 180 | 188 |
| treated | SSDKSVKVW | W | 9 | Q9GZS3 | WD repeat-containing protein 61 | 251 | 259 |
| treated | SSKPDPSQW | W | 9 | O75410 | Transforming acidic coiled-coil-containing protein 1 | 381 | 389 |
| treated | STKEMRNNW | W | 9 | Q63HN8 | RING finger protein 213 | 2742 | 2750 |
| treated | TTHNQSRKW | W | 9 | P47914 | 60S ribosomal protein L29 | 8 | 16 |
| treated | TTSKTDQRW | W | 9 | P49959 | Double-strand break repair protein MRE11A | 658 | 666 |
| treated | VSDASKPSW | W | 9 | Q7Z4G4 | tRNA guanosine-2'-O-methyltransferase TRM11 homolog | 288 | 296 |
| treated | VSEQHEYHW | W | 9 | Q7L590 | Protein MCM10 homolog | 798 | 806 |
| treated | VSNDRTQTW | W | 9 | Q9Y2H8 | Zinc finger protein 510 | 365 | 373 |
| treated | VSRILQEAW | W | 9 | Q9UI10 | Translation initiation factor eIF-2B subunit delta | 342 | 350 |
| treated | VTDNAVYHW | W | 9 | Q00610 | Clathrin heavy chain 1 | 122 | 130 |
| treated | VTNPHTDAW | W | 9 | O75293 | Growth arrest and DNA-damage-inducible protein GADD45 beta | 122 | 130 |
| treated | VVKNSVHSW | W | 9 | Q9NVI1 | Fanconi anemia group I protein | 367 | 375 |
| treated | AAKPEQIQKW | W | 10 | Q99986 | Serine/threonine-protein kinase VRK1 | 90 | 99 |
| treated | ATSTEKSVAW | W | 10 | Q8WZ64 | Centaurin-delta-1 | 311 | 320 |
| treated | AVNPRDPPSW | W | 10 | Q9Y490 | Talin-1 | 1621 | 1630 |
| treated | ESSRDASRKW | W | 10 | Q8TCU6 | Phosphatidylinositol 3,4,5-trisphosphate-dependent Rac exchanger | 1358 | 1367 |
| treated | GSADKTVALW | W | 10 | Q16576 | Histone-binding protein RBBP7 | 291 | 300 |
| treated | GSSDSTVRVW | W | 10 | Q9UKB1 | F-box/WD repeat-containing protein 11 | 296 | 305 |
| treated | HSVDSGSKRW | W | 10 | O75427 | Leucine-rich repeat/calponin homology domain-containing protein 4 | 303 | 312 |
| treated | HTTHNQSRKW | W | 10 | P47914 | 60S ribosomal protein L29 | 7 | 16 |
| treated | ISSSQDGHQW | W | 10 | P00451 | Coagulation factor VIII | 2281 | 2290 |
| treated | KALETDKKDW | W | 10 | O60645 | Exocyst complex component 3 | 408 | 417 |
| treated | KSRGSGEQDW | W | 10 | Q7Z5L9 | Interferon regulatory factor 2-binding protein 2 | 289 | 298 |
| treated | KSSIAGSSTW | W | 10 | Q99808 | Equilibrative nucleoside transporter 1 | 315 | 324 |
| treated | KSSTPNHKKW | W | 10 | P11388 | DNA topoisomerase 2-alpha | 599 | 608 |
| treated | KTGETASKRW | W | 10 | Q8TCF1 | AN1-type zinc finger protein 1 | 122 | 131 |
| treated | KVGSVVSVGW | W | 10 | Q8WUM9 | Sodium-dependent phosphate transporter 1 | 626 | 635 |
| treated | LALPPGAEHW | W | 10 | P42768 | Wiskott-Aldrich syndrome protein | 55 | 64 |
| treated | QANSHKERGW | W | 10 | P13051 | Uracil-DNA glycosylase | 222 | 231 |
| treated | RSRSTTAHSW | W | 10 | Q9UKJ3 | G patch domain-containing protein 8 | 560 | 569 |
| treated | RTMNIKSATW | W | 10 | Q16665 | Hypoxia-inducible factor 1 alpha | 180 | 189 |
| treated | RVRSHLAALW | W | 10 | P35613 | Basigin | 317 | 326 |
| treated | SATSSSQRDW | W | 10 | Q96T88 | E3 ubiquitin-protein ligase UHRF1 | 388 | 397 |
| treated | TSSSSVRVEW | W | 10 | Q9NZ45 | CDGSH iron sulfur domain-containing protein 1 | 4 | 13 |
| treated | VTQRKDDSTW | W | 10 | Q9HB58 | Sp110 nuclear body protein | 395 | 404 |
| treated | GTYSGKAADVW | W | 11 | Q96RU8 | Tribbles homolog 1 | 258 | 268 |
| treated | ITSQVTGQIGW | W | 11 | Q96CW1 | AP-2 complex subunit mu-1 | 151 | 161 |
| treated | KGDPQSPGRHW | W | 11 | Q5SYE7 | NHS-like protein 1 | 465 | 475 |
| treated | KTEDTQGKKKW | W | 11 | Q9ULH0 | Ankyrin repeat-rich membrane spanning protein | 641 | 651 |
| treated | LTKEDGEKETW | W | 11 | Q9NUA8 | Zinc finger and BTB domain-containing protein 40 | 670 | 680 |
| treated | QTKAKVTGDEW | W | 11 | P13807 | Glycogen [starch] synthase, muscle | 48 | 58 |
| treated | RSAPQGKGSSSW | W | 12 | Q99607 | ETS-related transcription factor Elf-4 | 337 | 348 |

| Sample | Peptide Sequence* | C-terminus | Length | Accession | Source Protein | Start | Stop |
|---|---|---|---|---|---|---|---|
| treated | RTPTKKEKDQAW | W | 12 | O75528 | Transcriptional adapter 3-like | 406 | 417 |
| treated | KSKTNSQKGASSTW | W | 14 | Q03701 | CCAAT/enhancer-binding protein zeta | 221 | 234 |
| treated | KTKQASKNTEKESAW | W | 15 | Q8WYP5 | AT-hook-containing transcription factor 1 | 2206 | 2220 |
| treated | RARESYKY | Y | 8 | Q9P0U4 | CpG-binding protein | 211 | 218 |
| treated | RTRKSINY | Y | 8 | Q9NRZ9 | Lymphoid-specific helicase | 511 | 518 |
| treated | ATSTIKRMY | Y | 9 | Q15361 | Transcription termination factor 1 | 509 | 517 |
| treated | ITKTVQQRY | Y | 9 | Q8IZT6 | Abnormal spindle-like microcephaly-associated protein | 2215 | 2223 |
| treated | KAAKLKEKY | Y | 9 | P09429 | High mobility group protein B1 | 147 | 155 |
| treated | KALIYQKKY | Y | 9 | Q99996 | A-kinase anchor protein 9 | 3717 | 3725 |
| treated | KSAVKNEEY | Y | 9 | Q14680 | Maternal embryonic leucine zipper kinase | 430 | 438 |
| treated | KSKVSEQRY | Y | 9 | Q9HA38 | Zinc finger matrin-type protein 3 | 271 | 279 |
| treated | KTGVDLRHY | Y | 9 | Q8N1B4 | Vacuolar protein sorting-associated protein 52 homolog | 75 | 83 |
| treated | KTKKIKEKY | Y | 9 | Q58FF8 | Putative heat shock protein HSP 90-beta 2 | 190 | 198 |
| treated | KTKKKYATY | Y | 9 | Q8IZT6 | Abnormal spindle-like microcephaly-associated protein | 2058 | 2066 |
| treated | KTNLRRTTY | Y | 9 | P17844 | Probable ATP-dependent RNA helicase DDX5 | 236 | 244 |
| treated | KTRGNTPKY | Y | 9 | O00567 | Nucleolar protein 5A | 340 | 348 |
| treated | KTVEPLEYY | Y | 9 | Q96B26 | Exosome complex exonuclease RRP43 | 6 | 14 |
| treated | KVRDQQLVY | Y | 9 | Q9UJX6 | Anaphase-promoting complex subunit 2 | 802 | 810 |
| treated | RSKEDGRLY | Y | 9 | Q99640 | Membrane-associated Tyr/Thr-specific cdc2-inhibitory kinase | 128 | 136 |
| treated | RSRsRDRMY | Y | 9 | Q9NYF8 | Bcl-2-associated transcription factor 1 | 45 | 53 |
| treated | RSRsREQSY | Y | 9 | Q96S94 | Cyclin-L2 | 383 | 391 |
| treated | RTIKKQRKY | Y | 9 | Q00059 | Transcription factor A, mitochondrial | 233 | 241 |
| treated | RVNTTQKRY | Y | 9 | P36578 | 60S ribosomal protein L4 | 114 | 122 |
| treated | ITRKGGERTY | Y | 10 | O75643 | U5 small nuclear ribonucleoprotein 200 kDa helicase | 596 | 605 |
| treated | KSKPVNKDRY | Y | 10 | P62316 | Small nuclear ribonucleoprotein Sm D2 | 86 | 95 |
| treated | KSLQNAQKHY | Y | 10 | Q14683 | Structural maintenance of chromosomes protein 1A | 317 | 326 |
| treated | KSRDGERTVY | Y | 10 | Q13263 | Transcription intermediary factor 1-beta | 199 | 208 |
| treated | KTKEGRDQLY | Y | 10 | Q9NS00 | Glycoprotein-N-acetylgalactosamine 3-beta-galactosyltransferase 1 | 135 | 144 |
| treated | RSKPPATVHY | Y | 10 | Q9NQC8 | UPF0360 protein C11orf60 | 191 | 200 |
| treated | RSMESRSHHY | Y | 10 | Q5THJ4 | Vacuolar protein sorting-associated protein 13D | 228 | 237 |
| treated | RSRDDADRRY | Y | 10 | Q5BJF6 | Outer dense fiber protein 2 | 763 | 772 |
| treated | RVADQRKKAY | Y | 10 | P09669 | Cytochrome c oxidase polypeptide VIc | 40 | 49 |
| treated | VTRPDSGHLY | Y | 10 | Q15418 | Ribosomal protein S6 kinase alpha-1 | 82 | 91 |
| treated | KAAEAHVDAHY | Y | 11 | P55263 | Adenosine kinase | 119 | 129 |
| treated | KTKPNTKDHGY | Y | 11 | Q12824 | SWI/SNF regulator of chromatin subfamily B member 1 | 71 | 81 |
| treated | KTYTGHKNEKY | Y | 11 | P61964 | WD repeat-containing protein 5 | 250 | 260 |
| untreated | PKKALHHF | F | 8 | O95602 | DNA-directed RNA polymerase I subunit RPA1 | 1062 | 1069 |
| untreated | RSKIAETF | F | 8 | Q9Y5A7 | NEDD8 ultimate buster 1 | 113 | 120 |
| untreated | RTRRAVKF | F | 8 | P83731 | 60S ribosomal protein L24 | 71 | 78 |
| untreated | ASLPLRVSF | F | 9 | O96005 | Cleft lip and palate transmembrane protein 1 | 299 | 307 |
| untreated | ATKKEITFF | F | 9 | P20592 | Interferon-induced GTP-binding protein Mx2 | 342 | 350 |
| untreated | ATRKAFTTF | F | 9 | Q14643 | Inositol 1,4,5-trisphosphate receptor type 1 | 1930 | 1938 |
| untreated | GTFRGTRYF | F | 9 | Q9NQC7 | Probable ubiquitin carboxyl-terminal hydrolase CYLD | 516 | 524 |
| untreated | GTKKQFQKF | F | 9 | P62280 | 40S ribosomal protein S11 | 150 | 158 |
| untreated | GVHTVHVTF | F | 9 | P21333 | Filamin-A | 452 | 460 |
| untreated | ISNPKTAEF | F | 9 | Q9NSD9 | Phenylalanyl-tRNA synthetase beta chain | 439 | 447 |
| untreated | ISVPIFKQF | F | 9 | Q00341 | Vigilin | 584 | 592 |
| untreated | ITKRKEVIF | F | 9 | Q8NCY6 | Coiled-coil domain-containing protein KIAA1826 | 24 | 32 |
| untreated | ITKTQKVRF | F | 9 | O60287 | Nucleolar pre-ribosomal-associated protein 1 | 257 | 265 |
| untreated | ITTTINPRF | F | 9 | P26010 | Integrin beta-7 | 782 | 790 |
| untreated | KAIEKNVLF | F | 9 | P10644 | cAMP-dependent protein kinase type I-alpha regulatory subunit | 130 | 138 |
| untreated | KAKDLILRF | F | 9 | Q15208 | Serine/threonine-protein kinase 38 | 352 | 360 |
| untreated | KAKDLLALF | F | 9 | Q9H501 | ESF1 homolog | 361 | 369 |
| untreated | KAKEIYMTF | F | 9 | O43665 | Regulator of G-protein signaling 10 | 81 | 89 |
| untreated | KALKRIHQF | F | 9 | O00311 | Cell division cycle 7-related protein kinase | 163 | 171 |
| untreated | KASGTIKKF | F | 9 | Q2NL67 | Poly [ADP-ribose] polymerase 6 | 162 | 170 |
| untreated | KASPTQNLF | F | 9 | P05120 | Plasminogen activator inhibitor 2 | 21 | 29 |
| untreated | KIKDFASHF | F | 9 | Q8ND82 | Zinc finger protein 280C | 627 | 635 |
| untreated | KSFDFHFGF | F | 9 | Q13432 | Protein unc-119 homolog A | 173 | 181 |
| untreated | KSHFRNHKF | F | 9 | Q5JTD0 | Tight junction-associated protein 1 | 183 | 191 |
| untreated | KSHTVVIGF | F | 9 | Q8NG31 | Protein CASC5 | 792 | 800 |
| untreated | KSKDISKTF | F | 9 | Q14651 | Plastin-1 | 84 | 92 |
| untreated | KSLEEIYLF | F | 9 | P15880 | 40S ribosomal protein S2 | 76 | 84 |

| Sample | Peptide Sequence* | C-terminus | Length | Accession | Source Protein | Start | Stop |
|---|---|---|---|---|---|---|---|
| untreated | KSMKVVKLF | F | 9 | Q9Y291 | 28S ribosomal protein S33, mitochondrial | 29 | 37 |
| untreated | KSVTLGYLF | F | 9 | P55265 | Double-stranded RNA-specific adenosine deaminase | 1063 | 1071 |
| untreated | KSWVFGVHF | F | 9 | P32248 | C-C chemokine receptor type 7 | 120 | 128 |
| untreated | KTAENFRQF | F | 9 | O43447 | Peptidyl-prolyl cis-trans isomerase H | 38 | 46 |
| untreated | KTFRPKRKF | F | 9 | Q86TA1 | Mps one binder kinase activator-like 2B | 13 | 21 |
| untreated | KTGNVVKRF | F | 9 | Q8NA23 | WD repeat-containing protein 31 | 91 | 99 |
| untreated | KTILPLINF | F | 9 | P16885 | 1-phosphatidylinositol-4,5-bisphosphate phosphodiesterase | 167 | 175 |
| untreated | KTLPDILTF | F | 9 | Q13569 | G/T mismatch-specific thymine DNA glycosylase | 122 | 130 |
| untreated | KVLDKMKSF | F | 9 | P06744 | Glucose-6-phosphate isomerase | 124 | 132 |
| untreated | LARRLLNRF | F | 9 | P46777 | 60S ribosomal protein L5 | 105 | 113 |
| untreated | LSKSDHSFF | F | 9 | P01906 | HLA class II histocompatibility antigen, DQ | 164 | 172 |
| untreated | LTAVVQKRF | F | 9 | P23396 | 40S ribosomal protein S3 | 69 | 77 |
| untreated | RAKAIIVEF | F | 9 | O60841 | Eukaryotic translation initiation factor 5B | 790 | 798 |
| untreated | RANVNARTF | F | 9 | Q00653 | Nuclear factor NF-kappa-B p100 subunit | 658 | 666 |
| untreated | RAYLAQRKF | F | 9 | O14579 | Coatomer subunit epsilon | 58 | 66 |
| untreated | RIKPRPERF | F | 9 | Q9NP77 | RNA polymerase II subunit A C-terminal domain phosphatase | 90 | 98 |
| untreated | RSFHEVREF | F | 9 | Q92963 | GTP-binding protein Rit1 | 106 | 114 |
| untreated | RSKPLFHHF | F | 9 | Q03113 | Guanine nucleotide-binding protein subunit alpha-12 | 342 | 350 |
| untreated | RSRQEPSRF | F | 9 | O14647 | Chromodomain-helicase-DNA-binding protein 2 | 114 | 122 |
| untreated | RSVLHHFHF | F | 9 | Q15269 | Periodic tryptophan protein 2 homolog | 87 | 95 |
| untreated | RTKPFHRKF | F | 9 | Q9Y5S2 | Serine/threonine-protein kinase MRCK beta | 1359 | 1367 |
| untreated | RTNIKRKTF | F | 9 | Q15022 | Polycomb protein SUZ12 | 124 | 132 |
| untreated | RTRRYVRKF | F | 9 | O43633 | Charged multivesicular body protein 2a | 68 | 76 |
| untreated | RTTNLIRHF | F | 9 | Q14592 | Zinc finger protein 460 | 459 | 467 |
| untreated | SSQKRHWTF | F | 9 | P51946 | Cyclin-H | 5 | 13 |
| untreated | TALPLLKQF | F | 9 | Q9Y2P8 | RNA 3'-terminal phosphate cyclase-like protein | 135 | 143 |
| untreated | VAKVGQYTF | F | 9 | Q86Y39 | NADH dehydrogenase [ubiquinone] 1 alpha subcomplex subunit 11 | 53 | 61 |
| untreated | VSAQIRKNF | F | 9 | Q8IU85 | Calcium/calmodulin-dependent protein kinase type 1D | 293 | 301 |
| untreated | VSKPDLITF | F | 9 | Q14586 | Zinc finger protein 267 | 47 | 55 |
| untreated | YAVRDREMF | F | 9 | O95298 | NADH dehydrogenase [ubiquinone] 1 subunit C2 | 82 | 90 |
| untreated | ISKQDAQELF | F | 10 | Q9BZJ0 | Crooked neck-like protein 1 | 443 | 452 |
| untreated | ITEKKAIKRF | F | 10 | O75643 | U5 small nuclear ribonucleoprotein 200 kDa helicase | 692 | 701 |
| untreated | KSAEHFKRKF | F | 10 | Q8IXW5 | RNA polymerase II-associated protein 2 | 332 | 341 |
| untreated | KSKESESFVF | F | 10 | A2VDJ0 | Transmembrane protein 131-like | 570 | 579 |
| untreated | KSLALKEKHF | F | 10 | Q92673 | Sortilin-related receptor | 2044 | 2053 |
| untreated | KSRGIGTVTF | F | 10 | P52272 | Heterogeneous nuclear ribonucleoprotein M | 242 | 251 |
| untreated | KSSQVQRRFF | F | 10 | P13010 | ATP-dependent DNA helicase 2 subunit 2 | 347 | 356 |
| untreated | KTRGIFETKF | F | 10 | P52789 | Hexokinase-2 | 777 | 786 |
| untreated | KTSEAKIKHF | F | 10 | Q02880 | DNA topoisomerase 2-beta | 221 | 230 |
| untreated | RAGPLSGKKF | F | 10 | P17844 | Probable ATP-dependent RNA helicase DDX5 | 25 | 34 |
| untreated | RSKEGLTERF | F | 10 | Q15046 | Lysyl-tRNA synthetase | 477 | 486 |
| untreated | RSVPHLQKVF | F | 10 | P07355 | Annexin A2 | 220 | 229 |
| untreated | TTDGYLLRLF | F | 10 | P61247 | 40S ribosomal protein S3a | 129 | 138 |
| untreated | ASKGPEALQEF | F | 11 | Q2TAL8 | Glutamine-rich protein 1 | 35 | 45 |
| untreated | ISAPDKRIYQF | F | 11 | O75563 | Src kinase-associated phosphoprotein 2 | 189 | 199 |
| untreated | ITSSAKVDMTF | F | 11 | Q14318 | FK506-binding protein 8 | 251 | 261 |
| untreated | KAFAETHIKGF | F | 11 | P42704 | Leucine-rich PPR motif-containing protein, mitochondrial | 1090 | 1100 |
| untreated | KARVETQNHWF | F | 11 | P28066 | Proteasome subunit alpha type-5 | 91 | 101 |
| untreated | KAVRDALKTEF | F | 11 | P56381 | ATP synthase subunit epsilon, mitochondrial | 21 | 31 |
| untreated | KSFETDTNLNF | F | 11 | P01848 | T-cell receptor alpha chain C region | 103 | 113 |
| untreated | KSKDIVNKMTF | F | 11 | P52732 | Kinesin-like protein KIF11 | 757 | 767 |
| untreated | KSRGSNLRVHF | F | 11 | P18621 | 60S ribosomal protein L17 | 16 | 26 |
| untreated | KTKTLDPKAVF | F | 11 | Q9HCE5 | Methyltransferase-like protein KIAA1627 | 271 | 281 |
| untreated | KTRGDHFKLRF | F | 11 | O43257 | Zinc finger HIT domain-containing protein 1 | 70 | 80 |
| untreated | KTRSSRAGLQF | F | 11 | Q96KK5 | Histone H2A type 1-H | 16 | 26 |
| untreated | KVDEVKSTIKF | F | 11 | P62906 | 60S ribosomal protein L10a | 147 | 157 |
| untreated | KVLKEIVERVF | F | 11 | Q14444 | Caprin-1 | 231 | 241 |
| untreated | RADPKKALHHF | F | 11 | O95602 | DNA-directed RNA polymerase I subunit RPA1 | 1059 | 1069 |
| untreated | RTKDLIIEQRF | F | 11 | Q00341 | Vigilin | 508 | 518 |
| untreated | RTLPVDSSHGF | F | 11 | Q6P4F7 | Rho GTPase-activating protein 11A | 328 | 338 |
| untreated | RVDPAKGLFYF | F | 11 | O75643 | U5 small nuclear ribonucleoprotein 200 kDa helicase | 666 | 676 |
| untreated | KSIDGKDAVHSF | F | 12 | Q8NEL9 | Phospholipase DDHD1 | 331 | 342 |
| untreated | KTKKPLLI | I | 8 | Q9BSR8 | Protein YIPF4 | 220 | 227 |

| Sample | Peptide Sequence* | C-terminus | Length | Accession | Source Protein | Start | Stop |
|---|---|---|---|---|---|---|---|
| untreated | FSKRIQKSI | I | 9 | O75446 | Histone deacetylase complex subunit SAP30 | 85 | 93 |
| untreated | HTKRMQHVI | I | 9 | Q9NV06 | WD repeat and SOF domain-containing protein 1 | 321 | 329 |
| untreated | KARDTKVLI | I | 9 | Q9UBH6 | Xenotropic and polytropic retrovirus receptor 1 | 679 | 687 |
| untreated | KSMKREYYI | I | 9 | Q8WYP5 | AT-hook-containing transcription factor 1 | 259 | 267 |
| untreated | KTKIYHPNI | I | 9 | P68036 | Ubiquitin-conjugating enzyme E2 L3 | 71 | 79 |
| untreated | KTRQIISTI | I | 9 | P37268 | Squalene synthetase | 358 | 366 |
| untreated | KTRQTIPRI | I | 9 | Q6SJ93 | Protein FAM111B | 337 | 345 |
| untreated | KVMSQRHMI | I | 9 | Q13148 | TAR DNA-binding protein 43 | 160 | 168 |
| untreated | RALPHHRVI | I | 9 | Q5JPE7 | Nodal modulator 2 | 1028 | 1036 |
| untreated | VSKPLAHHI | I | 9 | P55145 | Protein ARMET | 89 | 97 |
| untreated | KSAEAVGVKI | I | 10 | P27694 | Replication protein A 70 kDa DNA-binding subunit | 103 | 112 |
| untreated | KTKDHTLVQTI | I | 11 | Q14684 | Ribosomal RNA processing protein 1 homolog B | 202 | 212 |
| untreated | RTLEDRGIRKI | I | 11 | Q9Y2X3 | Nucleolar protein 5 | 396 | 406 |
| untreated | ASRPPVTL | L | 8 | Q15365 | Poly | 93 | 100 |
| untreated | ISSTVQRQL | L | 9 | Q6PJF5 | Rhomboid family member 2 | 386 | 394 |
| untreated | KALELERKL | L | 9 | Q9BWW8 | Apolipoprotein L6 | 274 | 282 |
| untreated | KAMELIREL | L | 9 | Q14691 | DNA replication complex GINS protein PSF1 | 5 | 13 |
| untreated | KAYEIEKRL | L | 9 | Q9UPP1 | PHD finger protein 8 | 381 | 389 |
| untreated | KAYPKRPLL | L | 9 | Q9GZR7 | ATP-dependent RNA helicase DDX24 | 381 | 389 |
| untreated | KSIKIVKKL | L | 9 | Q14692 | Ribosome biogenesis protein BMS1 homolog | 1019 | 1027 |
| untreated | KSIRQRNLL | L | 9 | P23458 | Tyrosine-protein kinase JAK1 | 227 | 235 |
| untreated | KTFTIKRFL | L | 9 | Q96EH5 | 60S ribosomal protein L39-like | 5 | 13 |
| untreated | KTFVGRAKL | L | 9 | Q9Y4K1 | Absent in melanoma 1 protein | 301 | 309 |
| untreated | KTKEAVLLL | L | 9 | P36578 | 60S ribosomal protein L4 | 163 | 171 |
| untreated | KTLATQRRL | L | 9 | Q9UIF9 | Bromodomain adjacent to zinc finger domain protein 2A | 772 | 780 |
| untreated | KTLERSYLL | L | 9 | P23921 | Ribonucleoside-diphosphate reductase large subunit | 149 | 157 |
| untreated | KTRSPKPLL | L | 9 | Q13724 | Mannosyl-oligosaccharide glucosidase | 108 | 116 |
| untreated | KTVRGRPKL | L | 9 | Q6NYC1 | Histone arginine demethylase JMJD6 | 300 | 308 |
| untreated | KVREIIQKL | L | 9 | P13797 | Plastin-3 | 49 | 57 |
| untreated | RSIEKRDTL | L | 9 | Q9H6F5 | Coiled-coil domain-containing protein 86 | 337 | 345 |
| untreated | RSRKLKLFL | L | 9 | P41252 | Isoleucyl-tRNA synthetase, cytoplasmic | 1225 | 1233 |
| untreated | RSVRLYHLL | L | 9 | Q14137 | Ribosome biogenesis protein BOP1 | 600 | 608 |
| untreated | RTVEKTWKL | L | 9 | Q13191 | E3 ubiquitin-protein ligase CBL-B | 44 | 52 |
| untreated | YSARQRRRL | L | 9 | P62841 | 40S ribosomal protein S15 | 37 | 45 |
| untreated | KISRETGEKL | L | 10 | O60880 | SH2 domain-containing protein 1A | 10 | 19 |
| untreated | KTMLDDDKRQL | L | 11 | Q9HAW4 | Claspin | 1104 | 1114 |
| untreated | VASKEIGKRKL | L | 11 | O60293 | Coiled-coil domain-containing protein 131 | 923 | 933 |
| untreated | WTSRQVGERTL | L | 11 | Q9C0K3 | Actin-related protein 11 | 24 | 34 |
| untreated | RTGKSYLM | M | 8 | Q96PP8 | Guanylate-binding protein 5 | 48 | 55 |
| untreated | KSIHIVVTM | M | 9 | Q13291 | Signaling lymphocytic activation molecule | 58 | 66 |
| untreated | KTKIDIIRM | M | 9 | Q16512 | Serine/threonine-protein kinase N1 | 174 | 182 |
| untreated | KTKVEVAKM | M | 9 | Q9NRD5 | PRKCA-binding protein | 81 | 89 |
| untreated | MARKINFLM | M | 9 | P82921 | 28S ribosomal protein S21, mitochondrial | 68 | 76 |
| untreated | KSLPAEINRM | M | 10 | Q9H9A6 | Leucine-rich repeat-containing protein 40 | 210 | 219 |
| untreated | HKQKGkVKRkLkNQ | Q | 14 | Q6UB98 | Ankyrin repeat domain-containing protein 12 | 466 | 479 |
| untreated | HSVEVHKW | W | 8 | Q96CP6 | GRAM domain-containing protein 1A | 674 | 681 |
| untreated | HSYRGHLW | W | 8 | P40337 | Von Hippel-Lindau disease tumor suppressor | 110 | 117 |
| untreated | IAKIPNFW | W | 8 | Q01105 | Protein SET | 88 | 95 |
| untreated | KIKVPVDW | W | 8 | P13796 | Plastin-2 | 432 | 439 |
| untreated | LSKIGEVW | W | 8 | Q8WYP5 | AT-hook-containing transcription factor 1 | 1068 | 1075 |
| untreated | RTRKEAAW | W | 8 | O15131 | Importin subunit alpha-6 | 393 | 400 |
| untreated | RVVKHFYW | W | 8 | O00626 | C-C motif chemokine 22 | 48 | 55 |
| untreated | VSKTSIGW | W | 8 | Q9NR09 | Baculoviral IAP repeat-containing protein 6 | 3295 | 3302 |
| untreated | AAKAVLADW | W | 9 | Q9NVN8 | Guanine nucleotide-binding protein-like 3-like protein | 384 | 392 |
| untreated | AARKHILVW | W | 9 | Q6P5X5 | UPF0545 protein C22orf39 | 77 | 85 |
| untreated | AARVLQEAW | W | 9 | O15554 | Intermed conductance calcium-activated potassium channel protein | 315 | 323 |
| untreated | ASAIIIQRW | W | 9 | Q8IZT6 | Abnormal spindle-like microcephaly-associated protein | 2068 | 2076 |
| untreated | ASAPLGARW | W | 9 | Q96AQ1 | Coiled-coil domain-containing protein 74A | 109 | 117 |
| untreated | ASKKVQRPW | W | 9 | P05166 | Propionyl-CoA carboxylase beta chain, mitochondrial | 523 | 531 |
| untreated | ASLDISRKW | W | 9 | Q9NWU5 | 39S ribosomal protein L22, mitochondrial | 37 | 45 |
| untreated | ATADVEWRW | W | 9 | Q96G21 | U3 small nucleolar ribonucleoprotein protein IMP4 | 267 | 275 |
| untreated | ATLKGNAAW | W | 9 | P07814 | Bifunctional aminoacyl-tRNA synthetase | 126 | 134 |
| untreated | ATLKNPILW | W | 9 | P04844 | Dolichyl-diphosphooligosaccharide--protein glycosyltransferase | 488 | 496 |

| Sample | Peptide Sequence* | C-terminus | Length | Accession | Source Protein | Start | Stop |
|---|---|---|---|---|---|---|---|
| untreated | ATRSGQDLW | W | 9 | Q9BV29 | Uncharacterized protein C15orf57 | 11 | 19 |
| untreated | ATVGRRYLW | W | 9 | P42166 | Lamina-associated polypeptide 2, isoform alpha | 646 | 654 |
| untreated | AVDSTVKVW | W | 9 | Q6Q0C0 | E3 ubiquitin-protein ligase TRAF7 | 660 | 668 |
| untreated | ETFHHSSNW | W | 9 | Q9UNY4 | Transcription termination factor 2 | 113 | 121 |
| untreated | FSLPAQPLW | W | 9 | O60216 | Double-strand-break repair protein rad21 homolog | 373 | 381 |
| untreated | GSASVNSRW | W | 9 | Q5SW79 | Centrosomal protein of 170 kDa | 1222 | 1230 |
| untreated | GSMGLRSLW | W | 9 | Q9BV87 | Uncharacterized protein C2orf24 | 301 | 309 |
| untreated | GTHSLDIKW | W | 9 | Q9UHQ1 | Nuclear prelamin A recognition factor | 448 | 456 |
| untreated | HAAKTKQAW | W | 9 | Q8N3C0 | Activating signal cointegrator 1 complex subunit 3 | 2019 | 2027 |
| untreated | HARPEEPSW | W | 9 | Q9NQ89 | Uncharacterized protein C12orf4 | 113 | 121 |
| untreated | HSLTKRREW | W | 9 | A6NDU8 | UPF0600 protein C5orf51 | 144 | 152 |
| untreated | HSRNLHHKW | W | 9 | Q13615 | Myotubularin-related protein 3 | 985 | 993 |
| untreated | HSSGILPKW | W | 9 | Q8NDX1 | PH and SEC7 domain-containing protein 4 | 475 | 483 |
| untreated | IAAQRTINW | W | 9 | Q5T4S7 | E3 ubiquitin-protein ligase UBR4 | 3282 | 3290 |
| untreated | IALPPIAKW | W | 9 | P50851 | Lipopolysaccharide-responsive and beige-like anchor protein | 217 | 225 |
| untreated | IAQSLEHSW | W | 9 | O43293 | Death-associated protein kinase 3 | 266 | 274 |
| untreated | ILRPPVEKW | W | 9 | Q5VYS8 | Zinc finger CCHC domain-containing protein 6 | 1432 | 1440 |
| untreated | IMKDKDNFW | W | 9 | P61160 | Actin-related protein 2 | 364 | 372 |
| untreated | ISAIPEQRW | W | 9 | O94953 | JmjC domain-containing histone demethylation protein 3B | 840 | 848 |
| untreated | ISGHHPETW | W | 9 | Q9Y385 | Ubiquitin-conjugating enzyme E2 J1 | 94 | 102 |
| untreated | IVRKDRHLW | W | 9 | Q8IZT6 | Abnormal spindle-like microcephaly-associated protein | 795 | 803 |
| untreated | KAKANLIGW | W | 9 | Q8WWQ0 | PH-interacting protein | 1813 | 1821 |
| untreated | KAKTNVKLW | W | 9 | P10644 | cAMP-dependent protein kinase type I-alpha regulatory subunit | 216 | 224 |
| untreated | KAVFLVPKW | W | 9 | Q9H2M9 | Rab3 GTPase-activating protein non-catalytic subunit | 96 | 104 |
| untreated | KGKNISSHW | W | 9 | Q96TA2 | ATP-dependent metalloprotease YME1L1 | 144 | 152 |
| untreated | KGRTEILKW | W | 9 | Q96TA2 | ATP-dependent metalloprotease YME1L1 | 510 | 518 |
| untreated | KILSNQEEW | W | 9 | O75791 | GRB2-related adapter protein 2 | 26 | 34 |
| untreated | KLKPLLEKW | W | 9 | Q9UKI9 | POU domain, class 2, transcription factor 3 | 244 | 252 |
| untreated | KLKTVKENW | W | 9 | Q00059 | Transcription factor A, mitochondrial | 181 | 189 |
| untreated | KSAFVRTQW | W | 9 | Q9UKY1 | Zinc fingers and homeoboxes protein 1 | 676 | 684 |
| untreated | KSGAIIEKW | W | 9 | Q67FW5 | UDP-GlcNAc:betaGal beta-1,3-N-acetylglucosaminyltransferase-like | 61 | 69 |
| untreated | KSIDLIQKW | W | 9 | P49585 | Choline-phosphate cytidylyltransferase A | 270 | 278 |
| untreated | KSRLSISGW | W | 9 | Q8N543 | 2-oxoglutarate/Fe-dependent oxygenase domain-containing protein | 228 | 236 |
| untreated | KTAEEKQKW | W | 9 | Q8TCU6 | Phosphatidylinositol 3,4,5-trisphosphate-dependent Rac exchanger | 375 | 383 |
| untreated | KTQGPRALW | W | 9 | Q96AG3 | Solute carrier family 25 member 46 | 152 | 160 |
| untreated | KTSSRRTTW | W | 9 | Q9BX40 | Protein LSM14 homolog B | 327 | 335 |
| untreated | KTSVVVGTW | W | 9 | Q96DZ1 | XTP3-transactivated gene B protein | 373 | 381 |
| untreated | KTVAIHTLW | W | 9 | Q92835 | Phosphatidylinositol-3,4,5-trisphosphate 5-phosphatase 1 | 478 | 486 |
| untreated | KVKHGHFGW | W | 9 | Q13501 | Sequestosome-1 | 187 | 195 |
| untreated | KVKKTIPSW | W | 9 | Q5SSJ5 | Heterochromatin protein 1-binding protein 3 | 128 | 136 |
| untreated | KVLLILSKW | W | 9 | Q9HCE0 | UPF0493 protein KIAA1632 | 2394 | 2402 |
| untreated | LARVLREDW | W | 9 | Q92845 | Kinesin-associated protein 3 | 179 | 187 |
| untreated | LSKKLVVRW | W | 9 | Q96H35 | Probable RNA-binding protein 18 | 95 | 103 |
| untreated | LSQPKIVKW | W | 9 | P61769 | Beta-2-microglobulin | 107 | 115 |
| untreated | LTSPDSEKW | W | 9 | Q9NR09 | Baculoviral IAP repeat-containing protein 6 | 670 | 678 |
| untreated | LTSRVTAHW | W | 9 | Q9UFC0 | Leucine-rich repeat and WD repeat-containing protein 1 | 141 | 149 |
| untreated | LVWKAQNTW | W | 9 | P02545 | Lamin-A/C | 512 | 520 |
| untreated | MSNKHGWTW | W | 9 | Q9Y3M8 | StAR-related lipid transfer protein 13 | 633 | 641 |
| untreated | NSRKQEAEW | W | 9 | P09496 | Clathrin light chain A | 127 | 135 |
| untreated | NSRSEAPNW | W | 9 | P60228 | Eukaryotic translation initiation factor 3 subunit E | 429 | 437 |
| untreated | QAFPNTNRW | W | 9 | P26641 | Elongation factor 1-gamma | 182 | 190 |
| untreated | QSAKEQIKW | W | 9 | P61923 | Coatomer subunit zeta-1 | 165 | 173 |
| untreated | QSHLNKALW | W | 9 | O60566 | Mitotic checkpoint serine/threonine-protein kinase BUB1 beta | 1028 | 1036 |
| untreated | QSLQIFRKW | W | 9 | Q5TA45 | Integrator complex subunit 11 | 332 | 340 |
| untreated | QTVSPAEKW | W | 9 | Q16658 | Fascin | 124 | 132 |
| untreated | RAGPKKESW | W | 9 | Q96GQ5 | UPF0420 protein C16orf58 | 409 | 417 |
| untreated | RAKDRKDVW | W | 9 | Q4LE39 | AT-rich interactive domain-containing protein 4B | 920 | 928 |
| untreated | RALAHYRWW | W | 9 | Q9Y2R9 | 28S ribosomal protein S7, mitochondrial | 234 | 242 |
| untreated | RALQLHLHW | W | 9 | Q16790 | Carbonic anhydrase 9 | 221 | 229 |
| untreated | RALVKRVTW | W | 9 | Q9P1Y6 | PHD and RING finger domain-containing protein 1 | 1397 | 1405 |
| untreated | RQSTILKRW | W | 9 | Q96CS7 | Pleckstrin homology domain-containing family B member 2 | 11 | 19 |
| untreated | RSHDTLVRW | W | 9 | Q9NU22 | Midasin | 177 | 185 |
| untreated | RTLRKDHRW | W | 9 | O14776 | Transcription elongation regulator 1 | 923 | 931 |

| Sample | Peptide Sequence* | C-terminus | Length | Accession | Source Protein | Start | Stop |
|---|---|---|---|---|---|---|---|
| untreated | RVAHFGYHW | W | 9 | O15239 | NADH dehydrogenase [ubiquinone] 1 alpha subcomplex subunit 1 | 37 | 45 |
| untreated | RVKTHLPSW | W | 9 | P17026 | Zinc finger protein 22 | 203 | 211 |
| untreated | RVLRERHLW | W | 9 | Q9Y3M8 | StAR-related lipid transfer protein 13 | 969 | 977 |
| untreated | RVRELYRAW | W | 9 | P56556 | NADH dehydrogenase [ubiquinone] 1 alpha subcomplex subunit 6 | 31 | 39 |
| untreated | RVRTPARQW | W | 9 | Q9BV79 | Trans-2-enoyl-CoA reductase, mitochondrial | 9 | 17 |
| untreated | SADGTLKLW | W | 9 | O14727 | Apoptotic protease-activating factor 1 | 762 | 770 |
| untreated | SAKRVAESW | W | 9 | Q9H7M6 | Zinc finger SWIM domain-containing protein 4 | 39 | 47 |
| untreated | SATRTLHEW | W | 9 | P27708 | CAD protein | 99 | 107 |
| untreated | SSKANPHRW | W | 9 | Q4G0I0 | Protein CCSMST1 | 54 | 62 |
| untreated | SSRNYQQHW | W | 9 | Q8TEM1 | Nuclear pore membrane glycoprotein 210 | 699 | 707 |
| untreated | SSRQIISHW | W | 9 | Q92793 | CREB-binding protein | 411 | 419 |
| untreated | SSVRAVAVW | W | 9 | Q9NNW5 | WD repeat-containing protein 6 | 775 | 783 |
| untreated | STDKTVRLW | W | 9 | O75529 | TAF5-like RNA polymerase II p300/CBP-associated factor | 445 | 453 |
| untreated | STDRHIRLW | W | 9 | Q9GZL7 | WD repeat-containing protein 12 | 316 | 324 |
| untreated | TAADVVKQW | W | 9 | Q15311 | RalA-binding protein 1 | 155 | 163 |
| untreated | TAKALQAHW | W | 9 | Q96EZ8 | Microspherule protein 1 | 247 | 255 |
| untreated | TSQDVLHSW | W | 9 | P00403 | Cytochrome c oxidase subunit 2 | 155 | 163 |
| untreated | TTAPEFRRW | W | 9 | P22695 | Cytochrome b-c1 complex subunit 2, mitochondrial | 141 | 149 |
| untreated | VAKKLGEMW | W | 9 | Q9UGV6 | High mobility group protein 1-like 10 | 125 | 133 |
| untreated | VSFAEKNGW | W | 9 | O43181 | NADH dehydrogenase iron-sulfur protein 4, mitochondrial | 135 | 143 |
| untreated | VSKIGDKNW | W | 9 | Q14008 | Cytoskeleton-associated protein 5 | 866 | 874 |
| untreated | VSTRQTQSW | W | 9 | Q96B23 | Uncharacterized protein C18orf25 | 365 | 373 |
| untreated | VTRAKQIVW | W | 9 | P00558 | Phosphoglycerate kinase 1 | 328 | 336 |
| untreated | YTKFQIATW | W | 9 | Q15800 | C-4 methylsterol oxidase | 44 | 52 |
| untreated | AAADSAVRLW | W | 10 | Q6IA86 | Elongator complex protein 2 | 133 | 142 |
| untreated | AAALPAAALW | W | 10 | Q03518 | Antigen peptide transporter 1 | 205 | 214 |
| untreated | ASARAGIHLW | W | 10 | Q9HAY2 | Melanoma-associated antigen F1 | 299 | 308 |
| untreated | ATILGNTERW | W | 10 | Q9UFH2 | Dynein heavy chain 17, axonemal | 3415 | 3424 |
| untreated | ATKVLGTVKW | W | 10 | P16989 | DNA-binding protein A | 88 | 97 |
| untreated | AVNKMDQVNW | W | 10 | Q9Y450 | HBS1-like protein | 404 | 413 |
| untreated | GSWDGTLRLW | W | 10 | P63244 | Guanine nucleotide-binding protein subunit beta-2-like 1 | 81 | 90 |
| untreated | GTKKYDLSKW | W | 10 | Q9UM54 | Myosin-VI | 1080 | 1089 |
| untreated | HSAKVHSVAW | W | 10 | Q96J01 | THO complex subunit 3 | 54 | 63 |
| untreated | HSARVGSLSW | W | 10 | Q12834 | Cell division cycle protein 20 homolog | 267 | 276 |
| untreated | HSLNKEARKW | W | 10 | O94868 | FCH and double SH3 domains protein 2 | 322 | 331 |
| untreated | HTADVQLHAW | W | 10 | Q8IWT0 | Protein archease | 36 | 45 |
| untreated | HTAVVEDVSW | W | 10 | Q09028 | Histone-binding protein RBBP4 | 226 | 235 |
| untreated | HTEARARHAW | W | 10 | P41273 | Tumor necrosis factor ligand superfamily member 9 | 217 | 226 |
| untreated | HTQDVKHVVW | W | 10 | O76071 | Protein CIAO1 | 149 | 158 |
| untreated | ISDPDVRHTW | W | 10 | P51805 | Plexin-A3 | 1706 | 1715 |
| untreated | KAFDEKKQKW | W | 10 | Q53T94 | TATA box-binding protein-associated factor RNA polymerase I | 409 | 418 |
| untreated | KAKHLASQYW | W | 10 | Q66K14 | TBC1 domain family member 9B | 828 | 837 |
| untreated | KALSDAIKKW | W | 10 | Q9NWF9 | E3 ubiquitin-protein ligase RNF216 | 406 | 415 |
| untreated | KARQGGDLGW | W | 10 | Q9Y237 | Peptidyl-prolyl cis-trans isomerase NIMA-interacting 4 | 75 | 84 |
| untreated | KAVDIVKQVW | W | 10 | Q460N5 | Poly [ADP-ribose] polymerase 14 | 645 | 654 |
| untreated | KAWRGTLARW | W | 10 | B0I1T2 | Myosin-Ig | 719 | 728 |
| untreated | KLKKPEAVRW | W | 10 | Q9Y2Z0 | Suppressor of G2 allele of SKP1 homolog | 245 | 254 |
| untreated | KLMPGRIQLW | W | 10 | A0AV96 | RNA-binding protein 47 | 213 | 222 |
| untreated | KSEALKDRHW | W | 10 | Q14204 | Cytoplasmic dynein 1 heavy chain 1 | 1404 | 1413 |
| untreated | KSFEKAKESW | W | 10 | Q13451 | FK506-binding protein 5 | 248 | 257 |
| untreated | KSHYDEAYKW | W | 10 | Q92624 | Amyloid protein-binding protein 2 | 219 | 228 |
| untreated | KSLSAERERW | W | 10 | Q14204 | Cytoplasmic dynein 1 heavy chain 1 | 3480 | 3489 |
| untreated | KSTPYTAVRW | W | 10 | O15258 | Protein RER1 | 35 | 44 |
| untreated | KSVKTGSVFW | W | 10 | Q8TCJ2 | Dolichyl-diphosphooligosaccharide--protein glycosyltransferase | 239 | 248 |
| untreated | KTKRVLPPNW | W | 10 | P62277 | 40S ribosomal protein S13 | 130 | 139 |
| untreated | KTLAEINQKW | W | 10 | Q99728 | BRCA1-associated RING domain protein 1 | 209 | 218 |
| untreated | KTPEKEPPLW | W | 10 | O14715 | RANBP2-like and GRIP domain-containing protein 8 | 626 | 635 |
| untreated | KTRDDWLVSW | W | 10 | P78527 | DNA-dependent protein kinase catalytic subunit | 3267 | 3276 |
| untreated | KTSELLVRKW | W | 10 | Q9BVM2 | Protein DPCD | 43 | 52 |
| untreated | KVFESWMHHW | W | 10 | P04233 | HLA class II histocompatibility antigen gamma chain | 179 | 188 |
| untreated | KVFKEKHHSW | W | 10 | Q5TBB1 | Ribonuclease H2 subunit B | 64 | 73 |
| untreated | KVREFNFEKW | W | 10 | P78347 | General transcription factor II-I | 345 | 354 |
| untreated | LTIKSIGHQW | W | 10 | P00403 | Cytochrome c oxidase subunit 2 | 95 | 104 |

| Sample | Peptide Sequence* | C-terminus | Length | Accession | Source Protein | Start | Stop |
|---|---|---|---|---|---|---|---|
| untreated | QTRRFQTETW | W | 10 | Q9UHB4 | NADPH-dependent diflavin oxidoreductase 1 | 587 | 596 |
| untreated | RARHAQGGTW | W | 10 | Q99832 | T-complex protein 1 subunit eta | 465 | 474 |
| untreated | RGFQSQVKKW | W | 10 | O43159 | Cerebral protein 1 | 282 | 291 |
| untreated | RGGDFKGRKW | W | 10 | Q96SK2 | Transmembrane protein 209 | 430 | 439 |
| untreated | RMYPLKPTWW | W | 10 | P05023 | Sodium/potassium-transporting ATPase subunit alpha-1 | 979 | 988 |
| untreated | RSKDVAKILW | W | 10 | Q7L8L6 | FAST kinase domain-containing protein 5 | 426 | 435 |
| untreated | RSLKDALFKW | W | 10 | P40818 | Ubiquitin carboxyl-terminal hydrolase 8 | 275 | 284 |
| untreated | RSNQWHGRSW | W | 10 | Q9BTL3 | Protein FAM103A1 | 78 | 87 |
| untreated | RSRARAGELW | W | 10 | Q9BXR0 | Queuine tRNA-ribosyltransferase | 9 | 18 |
| untreated | RSSQFARKLW | W | 10 | Q6ZNJ1 | Neurobeachin-like protein 2 | 2723 | 2732 |
| untreated | RSVAQAGVQW | W | 10 | Q6UX73 | Uncharacterized protein C16orf89 | 321 | 330 |
| untreated | RSVGRISKQW | W | 10 | Q9NRY6 | Phospholipid scramblase 3 | 231 | 240 |
| untreated | RSWDQQIKLW | W | 10 | Q14493 | Histone RNA hairpin-binding protein | 181 | 190 |
| untreated | RTAPFHLDLW | W | 10 | Q9NWW5 | Ceroid-lipofuscinosis neuronal protein 6 | 39 | 48 |
| untreated | RTAQLAKIKW | W | 10 | O00258 | Tryptophan-rich protein | 93 | 102 |
| untreated | RTRLEQVHEW | W | 10 | Q9P2E3 | NFX1-type zinc finger-containing protein 1 | 1731 | 1740 |
| untreated | RTYTYEKLLW | W | 10 | P35222 | Catenin beta-1 | 329 | 338 |
| untreated | RVDPDVAQHW | W | 10 | Q9NYR9 | NF-kappa-B inhibitor-interacting Ras-like protein 2 | 129 | 138 |
| untreated | RVINEEYKIW | W | 10 | Q16576 | Histone-binding protein RBBP7 | 14 | 23 |
| untreated | RVKAEPFIKW | W | 10 | P55010 | Eukaryotic translation initiation factor 5 | 372 | 381 |
| untreated | SAKSARLNLW | W | 10 | Q7KZF4 | Staphylococcal nuclease domain-containing protein 1 | 884 | 893 |
| untreated | SSKRAELEKW | W | 10 | Q9UH65 | Switch-associated protein 70 | 458 | 467 |
| untreated | SSSPEVKGYW | W | 10 | Q15366 | Poly | 270 | 279 |
| untreated | VAKPNIGENW | W | 10 | O60306 | Intron-binding protein aquarius | 522 | 531 |
| untreated | VSAEGVLHVW | W | 10 | P15260 | Interferon-gamma receptor alpha chain | 215 | 224 |
| untreated | VSMDEKKKEW | W | 10 | O15260 | Surfeit locus protein 4 | 260 | 269 |
| untreated | AALDSRKNYNW | W | 11 | Q9BYN8 | 28S ribosomal protein S26, mitochondrial | 179 | 189 |
| untreated | ATFHQRGIALW | W | 11 | P55884 | Eukaryotic translation initiation factor 3 subunit B | 350 | 360 |
| untreated | ATIKDIREHEW | W | 11 | P54646 | 5'-AMP-activated protein kinase catalytic subunit alpha-2 | 257 | 267 |
| untreated | FAYKDQNENRW | W | 11 | Q99590 | SFRS2-interacting protein | 772 | 782 |
| untreated | HSGLVHGLAFW | W | 11 | Q86X55 | Histone-arginine methyltransferase CARM1 | 380 | 390 |
| untreated | HSGPRGTHDLW | W | 11 | Q04446 | 1,4-alpha-glucan-branching enzyme | 312 | 322 |
| untreated | HSKENPKEFFW | W | 11 | Q8N999 | Uncharacterized protein C12orf29 | 95 | 105 |
| untreated | HSYSPRAIHSW | W | 11 | P28070 | Proteasome subunit beta type-4 | 126 | 136 |
| untreated | HTIGGSRRAAW | W | 11 | P61313 | 60S ribosomal protein L15 | 182 | 192 |
| untreated | ITIPDIKKDRW | W | 11 | O14757 | Serine/threonine-protein kinase Chk1 | 254 | 264 |
| untreated | ITYDDPIKTSW | W | 11 | Q9UJV9 | Probable ATP-dependent RNA helicase DDX41 | 136 | 146 |
| untreated | IVDPNGLARLW | W | 11 | P41212 | Transcription factor ETV6 | 370 | 380 |
| untreated | KAKYPDYEVTW | W | 11 | Q9NRX4 | 14 kDa phosphohistidine phosphatase | 110 | 120 |
| untreated | KITSVHSRIIW | W | 11 | Q6IA86 | Elongator complex protein 2 | 663 | 673 |
| untreated | KSFIQKKMQNW | W | 11 | Q9BY44 | Eukaryotic translation initiation factor 2A | 121 | 131 |
| untreated | KSFSKSDLVNW | W | 11 | O43847 | Nardilysin | 1060 | 1070 |
| untreated | KSIDAGPVDAW | W | 11 | Q9GZS3 | WD repeat-containing protein 61 | 100 | 110 |
| untreated | KSRDLFVSTSW | W | 11 | Q9BRQ8 | Apoptosis-inducing factor 2 | 355 | 365 |
| untreated | KSVSAAEQQLW | W | 11 | Q3B7T1 | Erythroid differentiation-related factor 1 | 866 | 876 |
| untreated | KTAGYQAVKRW | W | 11 | Q9P0U3 | Sentrin-specific protease 1 | 502 | 512 |
| untreated | KTVYTGIDHHW | W | 11 | Q9NV06 | WD repeat and SOF domain-containing protein 1 | 156 | 166 |
| untreated | KVDPAKDRELW | W | 11 | O60861 | Growth arrest-specific protein 7 | 450 | 460 |
| untreated | LTKQGGLVKTW | W | 11 | Q9UN19 | Dual Adapter for phosphotyrosine and 3-phosphoinositide | 171 | 181 |
| untreated | QTLKHPQTAKW | W | 11 | Q8WU76 | Sec1 family domain-containing protein 2 | 415 | 425 |
| untreated | RAAGALSKRYW | W | 11 | Q9H497 | Torsin-3A | 53 | 63 |
| untreated | RAFQAHKEENW | W | 11 | Q5JVF3 | PCI domain-containing protein 2 | 93 | 103 |
| untreated | RAIAAHERQAW | W | 11 | Q9NNW5 | WD repeat-containing protein 6 | 302 | 312 |
| untreated | RGTTILAKHAW | W | 11 | P51809 | Vesicle-associated membrane protein 7 | 10 | 20 |
| untreated | RSTLGHSRSHW | W | 11 | Q6ZNA4 | E3 ubiquitin-protein ligase Arkadia | 338 | 348 |
| untreated | RTREDIEGSHW | W | 11 | Q5XPI4 | E3 ubiquitin-protein ligase RNF123 | 715 | 725 |
| untreated | RVLPPSHRVTW | W | 11 | P15498 | Proto-oncogene vav | 15 | 25 |
| untreated | RVLPYPFTHHW | W | 11 | Q9BVJ6 | U3 small nucleolar RNA-associated protein 14 homolog A | 683 | 693 |
| untreated | SASRDKTIIMW | W | 11 | P63244 | Guanine nucleotide-binding protein subunit beta-2-like 1 | 33 | 43 |
| untreated | SSSLDAHIRLW | W | 11 | Q9GZS3 | WD repeat-containing protein 61 | 81 | 91 |
| untreated | SVSKDHALRLW | W | 11 | O75530 | Polycomb protein EED | 208 | 218 |
| untreated | TVSEIKASLKW | W | 11 | Q96H20 | Vacuolar-sorting protein SNF8 | 194 | 204 |
| untreated | VAKRNSLKELW | W | 11 | O43169 | Cytochrome b5 type B | 28 | 38 |

| Sample | Peptide Sequence* | C-terminus | Length | Accession | Source Protein | Start | Stop |
|---|---|---|---|---|---|---|---|
| untreated | VSFPDVEKAEW | W | 11 | Q9BSJ8 | Extended synaptotagmin-1 | 131 | 141 |
| untreated | VTLHDQGTAQW | W | 11 | P22314 | Ubiquitin-like modifier-activating enzyme 1 | 99 | 109 |
| untreated | VTSIGTAIRYW | W | 11 | O60337 | E3 ubiquitin-protein ligase MARCH6 | 83 | 93 |
| untreated | ASAGVDTNVRIW | W | 12 | Q13112 | Chromatin assembly factor 1 subunit B | 33 | 44 |
| untreated | ATSGNDGTIRVW | W | 12 | Q15751 | Probable E3 ubiquitin-protein ligase HERC1 | 3444 | 3455 |
| untreated | KLADPDEVARRW | W | 12 | P15036 | Protein C-ets-2 | 392 | 403 |
| untreated | KSFGDPAKPRAW | W | 12 | Q9Y4C8 | Probable RNA-binding protein 19 | 78 | 89 |
| untreated | KSHLHQKPGQTW | W | 12 | Q96Q83 | Alpha-ketoglutarate-dependent dioxygenase alkB homolog 3 | 31 | 42 |
| untreated | KSRGYVKEQFAW | W | 12 | P46783 | 40S ribosomal protein S10 | 53 | 64 |
| untreated | VSSSHDKSLRLW | W | 12 | Q9UNX4 | WD repeat-containing protein 3 | 691 | 702 |
| untreated | RTFGHSGIAVHTW | W | 13 | Q9P2Q2 | FERM domain-containing protein 4A | 285 | 297 |
| untreated | GTWKHARRY | Y | 9 | Q86T03 | Transmembrane protein 55B | 232 | 240 |
| untreated | ITRQKQLFY | Y | 9 | Q9Y690 | Putative transcription factor-like protein MORF4 | 87 | 95 |
| untreated | KARIQEAVY | Y | 9 | Q8TEM1 | Nuclear pore membrane glycoprotein 210 | 219 | 227 |
| untreated | KIRELQLRY | Y | 9 | Q5H9J7 | Protein BEX5 | 81 | 89 |
| untreated | KSFKKKFFY | Y | 9 | Q8N1G2 | FtsJ methyltransferase domain-containing protein 2 | 764 | 772 |
| untreated | KSFPGGKEY | Y | 9 | Q9Y6Q5 | AP-1 complex subunit mu-2 | 346 | 354 |
| untreated | KSLALLLNY | Y | 9 | Q9Y690 | Putative transcription factor-like protein MORF4 | 194 | 202 |
| untreated | KSRPELLEY | Y | 9 | Q14258 | Tripartite motif-containing protein 25 | 447 | 455 |
| untreated | KTRIPPRTY | Y | 9 | Q9BWU0 | Kanadaptin | 246 | 254 |
| untreated | LTVQVARVY | Y | 9 | P57076 | Uncharacterized protein C21orf59 | 27 | 35 |
| untreated | RIRRDVRVY | Y | 9 | Q86VU5 | Catechol-O-methyltransferase domain-containing protein 1 | 238 | 246 |
| untreated | RSILRNHRY | Y | 9 | Q6IN85 | Serine/threonine-protein phosphatase 4 regulatory subunit 3A | 660 | 668 |
| untreated | RSRDIRIKY | Y | 9 | O14511 | Pro-neuregulin-2, membrane-bound isoform | 279 | 287 |
| untreated | RTAHVILRY | Y | 9 | Q9UKV5 | Autocrine motility factor receptor, isoform 2 | 229 | 237 |
| untreated | RTKNNIQRY | Y | 9 | Q92665 | 28S ribosomal protein S31, mitochondrial | 51 | 59 |
| untreated | RTLPVLLLY | Y | 9 | Q9NR09 | Baculoviral IAP repeat-containing protein 6 | 597 | 605 |
| untreated | RTRVGVVRY | Y | 9 | Q8NFW1 | Collagen alpha-1 | 74 | 82 |
| untreated | RTVRIWRQY | Y | 9 | O76071 | Protein CIAO1 | 216 | 224 |
| untreated | VSFQHPHKY | Y | 9 | P0C7Q3 | Cyclin-related protein FAM58B | 140 | 148 |
| untreated | VTTVVNPKY | Y | 9 | P05556 | Integrin beta-1 | 787 | 795 |
| untreated | KARPIPRSTY | Y | 10 | Q3B820 | UPF0564 protein FAM161A | 356 | 365 |
| untreated | KSFPARLRQY | Y | 10 | Q14204 | Cytoplasmic dynein 1 heavy chain 1 | 1371 | 1380 |
| untreated | KSVENLGVSY | Y | 10 | P49643 | DNA primase large subunit | 54 | 63 |
| untreated | KTLTPIIQEY | Y | 10 | Q53EL6 | Programmed cell death protein 4 | 166 | 175 |
| untreated | KTVAGGAWTY | Y | 10 | P61513 | 60S ribosomal protein L37a | 62 | 71 |
| untreated | QTVKDSRTVY | Y | 10 | P78371 | T-complex protein 1 subunit beta | 399 | 408 |
| untreated | RSLSDLFRRY | Y | 10 | O75410 | Transforming acidic coiled-coil-containing protein 1 | 686 | 695 |
| untreated | RSRKESYSVY | Y | 10 | P62807 | Histone H2B type 1-C/E/F/G/I | 32 | 41 |
| untreated | ASMPRDIYQDY | Y | 11 | O15226 | NF-kappa-B-repressing factor | 90 | 100 |
| untreated | GSSLPADVHRY | Y | 11 | Q96KA5 | Cleft lip and palate transmembrane protein 1-like protein | 187 | 197 |
| untreated | ITKPQNLNDAY | Y | 11 | O60493 | Sorting nexin-3 | 12 | 22 |
| untreated | KIADMGHLKYY | Y | 11 | P12004 | Proliferating cell nuclear antigen | 240 | 250 |
| untreated | KSMGNLKHKQY | Y | 11 | Q96AP4 | Zinc finger with UFM1-specific peptidase domain protein | 538 | 548 |
| untreated | KSSPQIPHQTY | Y | 11 | Q16659 | Mitogen-activated protein kinase 6 | 703 | 713 |
| untreated | KTKPTHGIGKY | Y | 11 | Q96GC5 | 39S ribosomal protein L48, mitochondrial | 49 | 59 |
| untreated | KTLPADVQNYY | Y | 11 | Q99590 | SFRS2-interacting protein | 837 | 847 |
| untreated | KTMPNTTSRRY | Y | 11 | Q9UPV9 | Trafficking kinesin-binding protein 1 | 353 | 363 |
| untreated | KTRIISDGLKY | Y | 11 | P30876 | DNA-directed RNA polymerase II subunit RPB2 | 436 | 446 |
| untreated | KTRTLMEKDSY | Y | 11 | O15492 | Regulator of G-protein signaling 16 | 158 | 168 |
| untreated | KVRESERAFTY | Y | 11 | Q9BYT1 | Uncharacterized MFS-type transporter C20orf59 | 146 | 156 |
| untreated | RAKPEYISKTY | Y | 11 | Q13823 | Nucleolar GTP-binding protein 2 | 400 | 410 |
| untreated | KTLQHDKLVRLY | Y | 12 | P07948 | Tyrosine-protein kinase Lyn | 295 | 306 |

*s and t = pSer and pThr, respectively.